# Finding Cortical Subregions Regarding the Dorsal Language Pathway Based on the Structural Connectivity

Young-Eun Hwang[1,2,3], Young-Bo Kim[4,**] and Young-Don Son[2,3,*]

[1]Neuroscience Convergence Center, Korea University, Seoul 02841, Republic of Korea
[2]Department of Health Sciences and Technology, GAIHST, Gachon University, Incheon 21999, Republic of Korea
[3]Department of Biomedical Engineering, Gachon University, Incheon 21936, Republic of Korea
[4]Department of Neurosurgery, Gil Medical Center, College of Medicine, Gachon University, Incheon 21565, Republic of Korea

**\*Correspondence:**
**Young-Don Son Ph.D.**
Email: ydson@gachon.ac.kr
Address: 191 Hambakmoero, Yeonsu-gu, Incheon, 21936, Republic of Korea
Tel: +82 10-9119-4489

**\*\*Co-Correspondence:|**
**Young-Bo Kim, MD, Ph.D.**
E-mail: neurokim@gilhospital.com
Address: 21, Namdong-daero 774beon-gil, Namdong-gu, Incheon 21565, Republic of Korea
Tel: +82 10-7268-0550

## Abstract

Although the language-related fiber pathways in the human brain, such as the superior longitudinal fasciculus (SLF) and arcuate fasciculus (AF), are already well known, understanding more sophisticated cortical regions connected by the fiber tracts is essential to scrutinizing the structural connectivity of language circuits. With the regions of interest that were selected based on the Brainnetome atlas, the fiber orientation distribution estimation method for tractography was used to produce further elaborate connectivity information. The results indicated that both fiber bundles had two distinct connections with the prefrontal cortex (PFC). The SLF-II and dorsal AF are mainly connected to the rostrodorsal part of the inferior parietal cortex (IPC) and lateral part of the fusiform gyrus with the inferior frontal junction (IFJ), respectively. In contrast, the SLF-III and ventral AF were primarily linked to the anterior part of the supramarginal gyrus and superior part of the temporal cortex with the inferior frontal cortex, including the Broca's area. Moreover, the IFJ in the PFC, which has rarely been emphasized as a language-related subregion, also had the strongest connectivity with the previously known language-related subregions among the PFC; consequently, we proposed that these specific regions are interconnected via the SLF and AF within the PFC, IPC, and temporal cortex as language-related circuitry.

## 1. Introduction

Language is considered one of the most compelling indications of human cognitive activity. With much reasonable evidence that language activities proceed over macroscale networks in the human brain (Dick et al., 2012), finding language pathways has attracted significant interest from many researchers. As structural connectivity studies of neuronal fiber pathways increase rapidly, language researchers are focusing on constructing a consistent and systematic language framework that has been consolidated with lesion studies (Dick et al., 2012; Schmahmann et al., 2008; Bernal et al., 2009; Fridriksson et al., 2010). For example, abnormal or impaired fiber pathways may involve specific functional deficiency, so comprehending the language deficiency related to the specific lesion areas can offer crucial insight about clinical interaction. However, individual fiber bundles are difficult to identify, even though their names and functions have been studied separately. Above all, their spatial location, origin, and termination have been debated (Dick et al., 2012; Brauer et al., 2011; Yamada, 2009).

Among the well-known language pathways, the dorsal language stream, which controls phonological articulatory or word processing of speech (Chang et al., 2015; Saur et al., 2008; Yagmurlu et al., 2016), the superior longitudinal fasciculus (SLF), and the arcuate fasciculus (AF) in the left hemisphere, have been emphasized as the primary association fiber pathways crucial to language-related activity (Wernicke, 1908) and extensively studied over the past few decades. These association fiber bundles that connect the temporoparietal regions with the frontal areas were considered indissociable fiber bundles previously (Dick et al., 2012) but were separated from each other in recent studies. Moreover, there have been various efforts to subdivide SLF and AF further in recent years (Figure 1). Accordingly, the SLF, which connects the frontal cortex to the parietal cortex, has been generally divided into three branches, namely SLF-I, -II, and -III (Yagmurlu et al., 2016; Wang et al., 2015; Makris et al., 2005; Thiebaut de Schotten et al., 2011; Kamali et al., 2014), while the AF, which connects the frontal cortex and temporal lobe, has been broadly divided into two branches, namely dorsal and ventral AF (Glasser et al., 2008; Yagmurlu et al., 2016). Specifically, SLF-II and SLF-III are the pathways that link the inferior parietal lobule, otherwise known as Geschwind's territory, with the frontal regions and responsible for the articulatory aspects of language and visuospatial awareness (Wang et al., 2015; Yagmurlu et al., 2016; Fridriksson et al., 2010; Saur et al., 2008; Brownsett et sl., 2010; O'Connor et al., 2010; Wang et al., 2017). In contrast, the dorsal and ventral AF have been studied for their roles in phonological and semantic language processing (Poldrack et al., 1999; Keller et al., 2001; Yagmurlu et al., 2016). Moreover, even in terms of clinical correlation, the relative importance of the SLF and AF have been neglected compared with their cortical defects, but there is some definite evidence that patients with lesions of the SLF and AF had symptoms of conduction aphasia subtypes (Axer et al., 2001).

Growing language-related connectivity studies of neuronal fiber bundles have employed diverse diffusion-weighted magnetic resonance imaging (dMRI) (Le Bihan et al., 1985), which allowed the mapping of macrostructural connectivities in the human brain via the application of tractography or fiber tracking algorithms. Correspondingly, the language-related network of white matter connections, such as the frontoparietal and frontotemporal white matter tracts, has been demonstrated by various structural connectivity studies in vivo and compared with many functional imaging studies (Duffau et al., 2002, 2005, 2008; Hua et al., 2009; Powell et al., 2006, Saur et al., 2010).

However, the exact anatomical location and termination of the dorsal language pathway remain under debate (Dick et al., 2012; Brauer et al., 2011). Although consensus about SLF and AF division has been formed to some extent more recently (Table 1), slight differences persist regarding the exact cortical locations where the fiber tracts originate or terminate. In particular, the AF has been more disputed because the aspect of the fiber pathway is significantly different between human and non-human primates. In this regard, a recent study (Tremblay et al., 2016) reported that the current language model could not represent the distributed connectivity relevant to the language property due to use of outdated brain anatomy, which means that various subregions have not been thoroughly considered. Therefore, it is crucial to define subcomponents of language-related connectivity using improved brain models and dissociate their functional roles considering clinical interaction to identify more exact language-related regions.

In this study, we scrutinized the structural connectivity of the frontoparietal and frontotemporal fiber tracts in the left hemisphere using magnetic resonance (MR) diffusion tractography and the track termination distribution on the finer subregions defined by the Brainnetome atlas (Jiang, 2013; Fan et al., 2014, 2016). This study focuses on understanding the connectivity distribution of the dorsal language pathways and searching for anatomically more elaborate language-related cortical regions, specifically in the prefrontal cortex (PFC), inferior parietal cortex (IPC), and temporal cortex (TC).

## 2. Materials and Methods

### 2.1. Data set

In the Human Connectome Project (HCP) dataset (Van Essen et al., 2013), 1200 subjects were categorized by "HCP young adult" containing 7.0 T MRI data. Among the 1200 subjects, only 169 with dMRI were selected. The selected data were preprocessed with the HCP diffusion pipeline, which includes diffusion-weighting, diffusion-direction, time series, brain mask, log files of EDDY processing, and structurally extended preprocessed data. The preprocessing performed on the structural data was multimodal surface-matching registration for the diffusion processing pipeline. The subjects' ages ranged from 20 to 35 years, while the male-to-female ratio was 67–102.

For data acquisition for diffusion-weighted echo planar imaging, the following imaging parameters were used: voxel size = 1.05 mm3; frames = 71; TR/TE = 7000/71.2 ms; b-values = 1000 and 2000 s/mm2; and number of diffusion directions = 143. For an in-depth description of the data acquisition, processing, and analysis, see the HCP 1200 subjects release reference manual of the HCP (https://www.humanconnectome.org/storage/app/media/documentation/s1200/HCP_S1200_Release_Reference_Manual.pdf) (Van Essen et al., 2013; Sotiropoulos et al., 2013; Glasser et al., 2011; Andersson et al., 2003).

### 2.2. A series of data processing steps

During the data processing steps, the fiber orientation distribution (FOD) estimation method using the constrained spherical deconvolution (CSD) algorithm was utilized, which requires the acquiring of data using the high angular resolution diffusion-weighted imaging strategy (Tournier et al., 2004, 2007, Frank, 2001; Tuch et al., 2002). Tractography based on the FOD in each voxel and connectome process was performed using the MRrtrix3 software package (J-D Tournier, Brain Research Institute, Melbourne, Australia; https://github.com/MRtrix3/mrtrix3) (Tournier et al., 2012). To conduct a study that specifically observes cognitive and linguistic abilities, only the left hemisphere of the brain was considered. The workflow of the entire data processing step is shown in Figure 2.

### 2.2.1. Fiber orientation estimation

To estimate the full fiber orientation distribution function, CSD was performed on the MR signal data with a response function that describes a single coherently aligned white matter fiber population. This approach directly measures the signal orientation from the diffusion-weighted signal profile contained within each voxel. The obtained FOD contained orientation information and the corresponding volume fractions within each voxel (Tournier et al., 2004, 2007, 2008).

Recent studies have set the minimal value of the sampling density required for proper characterization of diffusion-weighted data. The signal profile measured in vivo contains significant information up to the spherical harmonics of order 8 (Tournier et al., 2011, 2013). Therefore, in this study, each FOD was estimated from the diffusion-weighted data using CSD algorithm with a lmax = 8.

### 2.2.2. Parcellation and Tractography

The preprocessed T1 images, more specifically segmented images of gray matter and white matter, were parcellated using the Brainnetome (http://atlas.brainnetome.org) (Jiang, 2013; Fan et al., 2014, 2016) and Desikan-Killiany (DK) atlas, respectively, using FreeSurfer software (http://surfer.nmr.mgh.harvard.edu/) (Fischi, 2012). The Brainnetome atlas was used to parcellate the cortical regions of interest, while the DK atlas was additionally used for white matter parcellation since the Brainnetome atlas provides only cortical parcellation information. Grey matter parcellation labels for the Brainnetome atlas were mapped to each subject using anatomical surface data. On the other hand, the white matter parcellation volume based on the DK atlas was generated in the segmentation steps by FreeSurfer.

This study mainly concentrated on the PFC, IPC, and TC regions. PFC was defined as Brodmann's area (BA) 8-14 and 44-47, which is located in the frontal portion of the frontal cortex (Murray et al., 2017). IPC was defined as BA 39 and 40, while TC was defined as BA 20-22, 28, 34-38, and 41-42. Based on the Brainnetome atlas, PFC, IPC, and TC were further divided into 23, 6, and 28 subregions, respectively.

To find the dorsal structural connectivity, such as the SLF and AF, the PFC, IPC, and TC were used as regions of interest for the tractography. Therefore, SLF (SLF-II and SLF-III) was set by originating from the IPC and terminating in the PFC, and AF was set by originating from the TC and terminating in the PFC. Finally, short fiber connectivity was generated in order to analyze the connectivity within the PFC.

With the parcellated gray matter and white matter information, probabilistic tractography was performed using the iFOD2 algorithm implemented in MRtrix3 to investigate the structural connectivity between PFC and IPC and between PFC and TC (Table 2, Figure 3). Relevant tracking parameters were: track minimum/maximum length = 60/180 mm; algorithm step size = 0.5 mm; curvature radius constraint = 0.8 mm; and FOD cutoff for track termination = 0.05. Streamline seeding was performed on white matter because the diffusion-weighted signal at the gray matter is relatively small, which could result in track distortion at the white matter–gray matter interface.

Furthermore, the endpoints of all the streamlines that cross the white matter–gray matter interface were precisely cropped, and anatomically constrained tractography was used to produce anatomically plausible streamlines that originate and terminate in the gray matter (Smith et al., 2012). The adequate values of step size and curvature constraint were determined through a parameter optimization process streamlined to the number of seeds. Additionally, to conduct a study that observes only linguistic abilities and their associated cognitive abilities, the right hemisphere of the brain was excluded, and irrelevant or unwanted cortexes were excluded considering the cortico-cortical connection property of the association fiber tracts.

### 2.2.3. Connectome

The connectivity matrix per subject was calculated based on the generated tractography streamlines and a node parcellation image that considered the region of interest of this study using the connectome algorithm implemented in MRtrix3 (Smith et al., 2015; Tournier et al., 2019). In this process, we set the option to perform a radial search from the endpoint of each streamline to locate the nearest node within 2 mm.

The parcellated T1 image was used to estimate the correlation matrix or track density matrix for the structural connectivity between subregions. These track density matrices were obtained by including the endpoints of each fiber on the cortical gray matter that matched the Brainnetome areas. Each track's density was calculated as the number of tracks reached to each subregion per voxel volume ratio of seeding subregions, which were randomly seeded and streamlined within a seed mask image. In other words, the calculated track density represents the connectivity ratio of tracks that reached the seeding subregions, which was set based only on the volume of the randomly seeded streamlines.

$$Trank\ Density = \frac{Number\ of\ reached\ tracks\ to\ each\ subregion}{\frac{Seeding\ subregion\ volume}{Seeding\ total\ volume} \times Number\ of\ total\ streamlines} \quad (eq.\ 1)$$

### 2.3. Data analysis

### 2.3.1. Finding major subregions connected via fiber tracts

Statistical analyses were performed using SPSS software (version 25.0; SPSS Inc., Chicago, IL, USA). The average track density matrix indicates the mean value of the calculated track density between the set subregions of 169 subjects. Based on this average track density matrix, the PFC subregions in which the number of reached tracks is in the 95th percentile were considered major subregions that have significant connectivity with the seeding areas. The number of tracks reaching the major PFC subregions was analyzed using box-and-whisker plots showing the distribution for the 169 subjects. Furthermore, a one-sample t-test was performed to determine which white matter connectivities are significantly different from the null hypothesis considering both sample size and data variability, which means that subregional connectivity has significant structural connectivity. In this process, any connectivity that had a value of $p < 0.05$ was considered statistically significant.

### 2.3.2 Three-dimensional visualization of structural connectivity

The average track density matrix was also visualized as a three-dimensional brain. In the three-dimensional visualization, each cortical subregion indicated the sum of the average track density related to that subregion and was represented by the relative density of a certain color across all regions of interest.

### 2.3.3. Mapping endpoints of streamlines to the IPC and TC

The opposite endpoints of the streamlines reaching the major PFC subregion were mapped to the seeding regions to confirm the connectivity distribution and find more specific anatomical connectivity locations within the seeding regions. For this process, the other fiber tracts, which were set to originate from each major PFC subregion, were generated. Relevant tracking parameters were set the same as the PFC-IPC and PFC-TC connectivities (Table 3).

To generate the track density maps of the streamline endpoints, track mapping algorithms implemented in MRtrix3 were utilized (Calamante et al., 2010). In this step, fiber tracts for each subject were registered to the Montreal Neurological Institute space and then combined for conversion into track density maps. The grid element on which the total number of tracks was calculated was set equal to the acquired voxel size, and only the endpoints were mapped to the image.

## 3. Results

### 3.1. Language-related connectivity in the human brain

#### 3.1.1. Connectivity between PFC and IPC

The average track density matrix for the structural connectivity between the PFC and IPC is shown in Figure 4(A). The matrix shows the mean value of the track density for all subjects. Among the PFC subregions, the inferior frontal junction (IFJ) and ventral area of BA44 (A44v) were considered the major PFC subregions (Supplementary Table 1). Figures 4(B) and 4(C) show the corresponding box-and-whisker plots showing the distribution of IFJ-IPC connectivity and A44v-IPC connectivity for all subjects, respectively. The matrix for the t-value comparison related to these subregional connectivities is shown in Figure 4(D). The result indicated that the connectivity between the IFJ and A40c (caudal area of BA40) had the highest t-value of 27.04. Moreover, IFJ had a noteworthy connectivity with A39rd (rostrodorsal area of BA39), A40rd (rostrodorsal area of BA40), and A39rv (rostroventral area of BA39), which showed t-values greater than 23. In contrast, the connectivity between A44v and A40rv (rostroventral area of BA40) had the second highest t-value of 26.39. Then, A44v had significant connectivity with A40c and A40rd, with t-values greater than 22 (Supplementary Table 2). According to these results, as we can see in the Figure 4(E), IFJ and A44v within the PFC had the strongest connectivity with the IPC, and anterior parts of the IPC had the strongest connectivity with the PFC (Supplementary Figure 1).

#### 3.1.2. Connectivity between PFC and TC

The average track density matrix between the PFC and TC is displayed in Figure 5(A), which indicates the mean track density for all subjects. Similar to the PFC-IPC connectivity, IFJ and A44v were considered the major PFC subregions (Supplementary Table 3), and Figures 5(B) and 5(C) show the box-and-whisker plot showing the distribution of IFJ-TC connectivity and A44v-TC connectivity for the 169 subjects, respectively. The corresponding matrix for the t-value comparison is shown in Figure 5(D). The results indicated that the white matter connectivity between the A44v and A22c (caudal area of BA22) was significantly correlated with the highest t-value of 24.95. A44v was also significantly connected with A41/42 (BA41 and 42) and TE1.0 & TE1.2 (anterior temporal visual association area), which showed t-values greater than 23. Furthermore, the A45c (caudal area of BA45) and IFS (inferior frontal sulcus) within the PFC had significant t-values with the A22c and A41/42 within the TC, which were greater than 21. In contrast, IFJ was significantly associated with A37dl (dorsolateral area of BA37), A37vl (ventrolateral area of BA37), and A37elv (extreme lateroventral area of BA37), which showed t-values great than 21 (Supplementary Table 4). Accordingly, as shown in Figure 5(E), the inferior frontal gyrus and the IFJ had significant connectivity with the TC, while the superior and posterior parts of the TC showed strong connectivity with the PFC (Supplementary Figure 2).

### 3.2. Fiber tract termination distribution on the IPC and TC

#### 3.2.1. Mapping IPC streamline endpoints

Figures 6(A) and (B) illustrate the IPC map for the streamline endpoints that originated from the IFJ or A44v, respectively. These show a more specific anatomical location and topological analysis of the structural connectivity. The endpoint density of the streamlines originating from each major PFC subregion differed. The IFJ had strong connectivity with the superior parts of the IPC (A40c, A40rd, and A39rd), while the A44v, otherwise known as Broca's area, had strong connectivity with the anterior part of the supramarginal gyrus (A40rv, A40rd, and A40c). This allowed us to obtain more accurate cortical regions for each major fiber connectivity. Accordingly, this could be a further result of previous studies of SLF-II and SLF-III.

#### 3.2.2. Mapping TC streamline endpoints

Figure 7 represents the TC mapping of the endpoints of the streamlines that connect the TC with the IFJ or A44v. In the case of TC, the cortical subregions linked with IFJ and A44v were separated more clearly. In summary, BA37, also known as the fusiform gyrus, has significant connectivity with the IFJ, while the A44v (part of Broca's area) shows strong connectivity with A41/42, which is known as the auditory cortex, and A22c (otherwise known as Wernicke's area). This finding is consistent with previous study results of the AF being divided into dorsal and ventral portions and provides more specific anatomical locations.

### 3.3. Short fiber connectivity within the PFC

Overall, the connectivity results indicated that IFJ and A44v had the strongest structural connectivity in the SLF and AF fiber bundles, indicating that these subregions could play a vital role in language tasks. Therefore, we assume that IFJ and A44v also have considerable connectivity with language-related regions within the PFC.

To analyze the connectivity within the PFC, a short fiber connectivity matrix was produced by applying the same tractography process to the PFC area. As shown in Figure 8(A), the resulting short fiber connectivity matrix shows that A44v has the highest mean value with A8dl (dorsolateral area of BA8), while IFJ has relatively strong connectivity with A9/46v (ventral area of BA 9 and BA 46) areas. Figures 8(B) and (C) show a box-and-whisker plot of the distribution of the IFJ and A44v connectivities indicated in the average track density matrix. The corresponding matrix for the t-value comparison is shown in Figure 8(D). The fiber tract connectivity linking the IFJ and A9/46v had a noticeable t-value compared to other connectivities of 27.85 (Figure 8(B), Supplementary Table 5). On the other hand, A44v had significant connectivity with A9/46v and A45c with t-values greater than 21. Considering the average track density and significance, IFJ had significant connectivity with A9/46v within the PFC (Figure 8(E), Supplementary Figure 3).

### 4. Discussions

In the present study, brain regions that were interconnected via the SLF and AF were examined and further specified along with various previous studies. As shown in Figures 9(A) and (B), the structural connectivity that links the PFC with the IPC (i.e. the SLF-II and SLF-III) and the TC (i.e. the AF) showed that the IFJ and A44v of the PFC had significant connectivity. Furthermore, the rostrodorsal region of the IPC and the lateral part of the fusiform gyrus had the strongest connection to the IFJ, while the anterior part of the supramarginal gyrus and the superior regions of the TC had significant connectivity with the A44v.

To be more concrete, we specified the middle frontal gyrus as the IFJ, which had the major white matter connectivity with parts of the SLF-III and the dorsal AF. Among the previous connectivity studies of the language pathway, there is consensus that the SLF-III and the dorsal AF are connected to the middle frontal gyrus (Yagmurlu et al., 2016; Wang et al., 2015; Makris et al., 2005; Thiebaut de Schotten et al., 2011; Kamali et al., 2014; Glasser et al., 2008). Based on our results, we confirmed that IFJ within the middle frontal gyrus had significant connectivity with the IPC and the TC. A previous cognitive study identified that the IFJ area lies at the junction of the premotor and the prefrontal cortex as part of the mid-dorsolateral prefrontal cortex (DLPFC), which is located at the junction of the premotor, language, and working memory domains (Brass et al., 2005). Moreover, there has been much speculation that the IFJ is ideally located to promote the interaction of information among these domains based on various functional imaging studies (Derrfuss et al., 2004, 2005; Neumann et al., 2005; Brass et al., 2005). In this respect, the IFJ within the PFC might serve a crucial role in language processing by communicating with the IPC and the TC.

On the other hand, the inferior frontal gyrus, which was interconnected with the IPC and the TC via the SLF-III and the ventral AF, was further concretized as the ventral part of the pars opercularis (A44v). According to the previous studies, the SLF-III and the ventral AF are connected to the inferior frontal gyrus (Yagmurlu et al., 2016; Wang et al., 2015; Makris et al., 2005; Thiebaut de Schotten et al., 2011; Kamali et al., 2014; Glasser et al., 2008). Based on our results, the inferior frontal gyrus had strong structural connectivity; however, the ventral area of BA44 had much more direct connectivity with the IPC and the TC than other regions related to the inferior frontal gyrus. It is well known that the pars opercularis (BA44) and pars triangularis (BA45) are associated with Broca's area and phonological rehearsal (Na et al., 2000; Baldo et al., 2006; Smith et al., 1988). Although there has been no clear functional separation between BA44 and BA45, A44v is expected to play an important role in the language circuitry.

Based on these major PFC subregions in our study, we investigated a more specific cortical region of IPC in which each fiber tract branch was interconnected. The superior portion (A39rd, A40c, and A40rd) and anterior parts (A40rv, A40rd, and A40c) of the IPC had the strongest connectivity with the IFJ and A44v, respectively. In this context, SLF-II and SLF-III are known to originate from the angular gyrus and the supramarginal gyrus, respectively (Yagmurlu et al., 2016; Wang et al., 2015; Makris et al., 2005; Thiebaut de Schotten et al., 2011; Kamali et al., 2014). First, the superior portion of the IPC (A39rd, A40c and A40rd) had the strongest connectivity with the IFJ. This connectivity could be considered the SLF-II; however, it was slightly different from previous studies in which it originated from the superior parts of the whole IPC rather than from the angular gyrus only. A previous subregional study (Ding et al., 2020) reported that damage to the rostrodorsal regions of the IPL resulted in impaired lexical selection and reduced structural complexity for speech production.

In contrast, the anterior parts of the IPC (A40rv, A40rd, and A40c) had significant connectivity with A44v via the SLF-III. This connectivity finding was consistent with those of previous studies in that it originated from the supramarginal gyrus (Yagmurlu et al., 2016; Wang et al., 2015; Makris et al., 2005; Thiebaut de Schotten et al., 2011; Kamali et al., 2014). Furthermore, we identified that it was intensively interconnected with the anterior parts of the supramarginal gyrus. According to the dissociation analysis, the rostral supramarginal gyrus was significantly correlated with syntactic accuracy deficits along with the pars opercularis (Ding et al., 2020). Regarding these results, we believe that the SLF-II and SLF-III, which could be dissociated by specific originated regions, reflected different functions for grammatical word production and syntactic accuracy.

Similar to the SLF, more specific regions of the TC, in which each AF branch was interconnected, were investigated based on the major PFC subregions. For the dorsal AF, the dorsolateral area of the fusiform gyrus (A37dl) within the TC showed the strongest connectivity with the IFJ. This AF branch originated from the posterior parts of the middle and inferior temporal gyrus (Yagmurlu et al., 2016; Glasser et al., 2008). This result was in accordance with those of previous studies and provided further specific regions. The fusiform gyrus (BA37) region was previously studied for its role in visual perception and semantic language function (Ardila et al., 2015; Jouen et al., 2015; Ding et al., 2020). In this regard, it was in line with the results of previous studies (Glasser et al., 2008; Yagmurlu et al., 2016; Ding et al., 2020) that the dorsal AF could be related to the semantic language pathway.

For the ventral AF, the posterior superior temporal gyri (A41/42 and A22c), well-known interconnected regions via the classical AF pathway, showed strong direct connectivity with A44v. This result was consistent with that of the original AF study in which the AF originated from the superior temporal gyrus (Yagmurlu et al., 2016; Glasser et al., 2008). Lesions in the posterior superior temporal gyrus could reportedly lead to abnormalities in phonological retrieval and speech articulation (Binder, 2015). Specifically, A22c, known as Wernicke's area, has been discussed as a region related to language comprehension abnormalities (Binder, 2015). Based on previous studies (Binder, 2015; Barbey et al., 2013; Na et al., 2000; Baldo et al., 2006; Smith et al., 1988) and the current results, the connectivity between A44v and the posterior superior temporal gyrus might play a crucial role in phonological pathways that convert the stimulus to a phonological form.

Finally, realizing the importance of the IFJ and A44v, short fiber connectivity within the PFC was also observed. The resulting data showed that the IFJ has the strongest connectivity with A9/46v (otherwise known as the DLPFC) (Figure 9(C)), which is known to activate the central executive system related to verbal working memory (Townsend et al., 2010; Snow et al., 2016; Barbey et al., 2013; Na et al., 2000; Baldo et al., 2006; Smith et al., 1988). Previous studies (Brass et al., 2005) proved the important relationship between working memory, language comprehension, and cognition. These studies provided evidence on the role of the frontoparietal and frontotemporal pathways in assisting language tasks that require working memory, such as in a verbal working memory task or a repetition task that reproduces the comprehension of auditorily perceived words (Brass et al., 2005; Derrfuss et al., 2005). Therefore, we hypothesize that the IFJ may play a crucial role in supporting complex processing through some form of working memory-related mechanism as an entrance of the PFC in the language pathway.

Despite our attempt to dissociate further specific regions within the IFJ and A44v where each fiber tract (SLF and AF, respectively) reached, we could not confirm a more specific endpoint distribution of each connectivity because of the limited MRI resolution and the probabilistic property of tractography. In the current study, each tract that was terminated to the IFJ (SLF-II and dorsal AF) and A44v (SLF-III and ventral AF) was indistinguishable because they might have merged with each other while extending to the frontal lobe. Accordingly, the cortical regions could be defined more accurately in future studies using submillimeter MRI data.

## 5. Conclusions

The present study demonstrated the subregional structure connectivity of the SLF and AF tracts in the context of the language pathway. The fiber track density within the subregions was calculated using MR diffusion tractography. Both fiber bundles had two main aspects. First, the rostrodorsal regions of the IPC and the lateral parts of the fusiform gyrus had the strongest connectivity with the IFJ within the PFC via the SLF-III and dorsal AF. Second, the anterior parts of the supramarginal gyrus and the superior region of the TC had significant connectivity with A44v within the PFC via the SLF-II and ventral AF, respectively. The IFJ also showed the strongest connectivity with the DLPFC among the rest of the PFC regions via short fiber tracts. In conclusion, this study emphasized specific language-related regions of the human brain to further understand the language circuitry.

**Declarations**

**Data availability statement**

Data were provided [in part] by the Human Connectome Project, WU-Minn Consortium (Principal Investigators: David Van Essen and Kamil Ugurbil; 1U54MH091657) funded by the 16 NIH Institutes and Centers that support the NIH Blueprint for Neuroscience Research; and by the McDonnell Center for Systems Neuroscience at Washington University.

For in-depth description on data acquisition, processing and analysis, see the HCP 1200 Subjects Release Reference Manual of the HCP (https://www.humanconnectome.org/storage/app/media/documentation/s1200/HCP_S1200_Release_Reference_Manual.pdf ).

**Funding Statement**

This work was supported by the Brain Research Program of the National Research Foundation of Korea (NRF), which is funded by the Ministry of Science and ICT (2017M3C7A1049026) and by the National Research Foundation of Korea (NRF) grant funded by the Korea government (MSIT) (No. NRF-2020R1A4A1019623).

**Conflict of interest disclosure**

The authors declare that they have no competing interest.

**Acknowledgement**

This work was supported by the Brain Research Program of the National Research Foundation of Korea (NRF), which is funded by the Ministry of Science and ICT (2017M3C7A1049026) and by the National Research Foundation of Korea (NRF) grant funded by the Korea government (MSIT) (No. NRF-2020R1A4A1019623).

Data were provided [in part] by the Human Connectome Project, WU-Minn Consortium (Principal Investigators: David Van Essen and Kamil Ugurbil; 1U54MH091657) funded by the 16 NIH Institutes and Centers that support the NIH Blueprint for Neuroscience Research; and by the McDonnell Center for Systems Neuroscience at Washington University.

**Author's contributions**

Young-Eun Hwang contributed to the study design, data analysis and interpretation, and manuscript writing. Young-Don Son and Young-Bo Kim contributed to the study design, directed the project, data interpretation and critically revised the manuscript. All authors gave their final approval and agreed to be accountable for all aspects of the work.

**References**

[1] Dick AS, Tremblay P. Beyond the arcuate fasciculus: consensus and controversy in the connectional anatomy of language. Brain. 2012 Dec;135(Pt 12):3529-50. doi: 10.1093/brain/aws222. Epub 2012 Oct 29. PMID: 23107648.

[2] Schmahmann JD, Smith EE, Eichler FS, Filley CM. Cerebral white matter: neuroanatomy, clinical neurology, and neurobehavioral correlates. Ann N Y Acad Sci. 2008 Oct;1142:266-309. doi: 10.1196/annals.1444.017. PMID: 18990132; PMCID: PMC3753195.

[3] Bernal B, Ardila A. The role of the arcuate fasciculus in conduction aphasia. Brain. 2009 Sep;132(Pt 9):2309-16. doi: 10.1093/brain/awp206. Epub 2009 Aug 18. PMID: 19690094.

[4] Fridriksson J, Kjartansson O, Morgan PS, Hjaltason H, Magnusdottir S, Bonilha L, Rorden C. Impaired speech repetition and left parietal lobe damage. J Neurosci. 2010 Aug 18;30(33):11057-61. doi: 10.1523/JNEUROSCI.1120-10.2010. PMID: 20720112; PMCID: PMC2936270.

[5] Chang EF, Raygor KP, Berger MS. Contemporary model of language organization: an overview for neurosurgeons. J Neurosurg. 2015 Feb;122(2):250-61. doi: 10.3171/2014.10.JNS132647. Epub 2014 Nov 28. PMID: 25423277.

[6] Saur D, Kreher BW, Schnell S, Kümmerer D, Kellmeyer P, Vry MS, Umarova R, Musso M, Glauche V, Abel S, Huber W, Rijntjes M, Hennig J, Weiller C. Ventral and dorsal pathways for language. Proc Natl Acad Sci U S A. 2008 Nov 18;105(46):18035-40. doi: 10.1073/pnas.0805234105. Epub 2008 Nov 12. PMID: 19004769; PMCID: PMC2584675.

[7] Kamali A, Flanders AE, Brody J, Hunter JV, Hasan KM. Tracing superior longitudinal fasciculus connectivity in the human brain using high resolution diffusion tensor tractography. Brain Struct Funct. 2014 Jan;219(1):269-81. doi: 10.1007/s00429-012-0498-y. Epub 2013 Jan 4. PMID: 23288254; PMCID: PMC3633629.

[8] Yagmurlu K, Middlebrooks EH, Tanriover N, Rhoton AL Jr. Fiber tracts of the dorsal language stream in the human brain. J Neurosurg. 2016 May;124(5):1396-405. doi: 10.3171/2015.5.JNS15455. Epub 2015 Nov 20. PMID: 26587654.

[9] Wernicke C. The symptom-complex of aphasia, Diseases of the nervous system, 1908 New York Appleton (pg. 265-324)

[10] Wang X, Pathak S, Stefaneanu L, Yeh FC, Li S, Fernandez-Miranda JC. Subcomponents and connectivity of the superior longitudinal fasciculus in the human brain. Brain Struct Funct. 2016 May;221(4):2075-92. doi: 10.1007/s00429-015-1028-5. Epub 2015 Mar 18. PMID: 25782434.

[11] Makris N, Kennedy DN, McInerney S, Sorensen AG, Wang R, Caviness VS Jr, Pandya DN. Segmentation of subcomponents within the superior longitudinal fascicle in humans: a quantitative, in vivo, DT-MRI study. Cereb Cortex. 2005 Jun;15(6):854-69. doi: 10.1093/cercor/bhh186. Epub 2004 Dec 8. PMID: 15590909.

[12] Catani M, Jones DK, ffytche DH. Perisylvian language networks of the human brain. Ann Neurol. 2005 Jan;57(1):8-16. doi: 10.1002/ana.20319. PMID: 15597383.

[13] Thiebaut de Schotten M, Dell'Acqua F, Forkel SJ, Simmons A, Vergani F, Murphy DG, Catani M. A lateralized brain network for visuospatial attention. Nat Neurosci. 2011 Sep 18;14(10):1245-6. doi: 10.1038/nn.2905. Erratum in: Nat Neurosci. 2011 Dec;14(12):1617. PMID: 21926985.

[14] Glasser MF, Rilling JK. DTI tractography of the human brain's language pathways. Cereb Cortex. 2008 Nov;18(11):2471-82. doi: 10.1093/cercor/bhn011. Epub 2008 Feb 14. PMID: 18281301.

[15] Brownsett SL, Wise RJ. The contribution of the parietal lobes to speaking and writing. Cereb Cortex. 2010 Mar;20(3):517-23. doi: 10.1093/cercor/bhp120. Epub 2009 Jun 16. PMID: 19531538; PMCID: PMC2820696.

[16] O'Connor AR, Han S, Dobbins IG. The inferior parietal lobule and recognition memory: expectancy violation or successful retrieval? J Neurosci. 2010 Feb 24;30(8):2924-34. doi: 10.1523/JNEUROSCI.4225-09.2010. PMID: 20181590; PMCID: PMC2844718.

[17] Wang, M., Zhang, J., Dong, G., Zhang, H., Lu, H. and Du, X. (2017), Development of rostral inferior parietal lobule area functional connectivity from late childhood to early adulthood. International Journal of Developmental Neuroscience, 59: 31-36. doi:10.1016/j.ijdevneu.2017.03.001

[18] Poldrack RA, Wagner AD, Prull MW, Desmond JE, Glover GH, Gabrieli JD. Functional specialization for semantic and phonological processing in the left inferior prefrontal cortex. Neuroimage. 1999 Jul;10(1):15-35. doi: 10.1006/nimg.1999.0441. PMID: 10385578.

[19] Keller TA, Carpenter PA, Just MA. The neural bases of sentence comprehension: a fMRI examination of syntactic and lexical processing. Cereb Cortex. 2001 Mar;11(3):223-37. doi: 10.1093/cercor/11.3.223. PMID: 11230094.

[20] Le Bihan D, Breton E. Imagerie de diffusion in-vivo par résonance magnétique nucléaire. C R Acad Sci. 1985;301:1109-1112.

[21] Duffau H, Capelle L, Sichez N, Denvil D, Lopes M, Sichez JP, Bitar A, Fohanno D. Intraoperative mapping of the subcortical language pathways using direct stimulations. An anatomo-functional study. Brain. 2002 Jan;125(Pt 1):199-214. doi: 10.1093/brain/awf016. PMID: 11834604.

[22] Duffau H, Gatignol P, Mandonnet E, Peruzzi P, Tzourio-Mazoyer N, Capelle L. New insights into the anatomo-functional connectivity of the semantic system: a study using cortico-subcortical electrostimulations. Brain. 2005 Apr;128(Pt 4):797-810. doi: 10.1093/brain/awh423. Epub 2005 Feb 10. PMID: 15705610.

[23] Duffau H. The anatomo-functional connectivity of language revisited. New insights provided by electrostimulation and tractography. Neuropsychologia. 2008 Mar 7;46(4):927-34. doi: 10.1016/j.neuropsychologia.2007.10.025. Epub 2007 Nov 17. PMID: 18093622.

[24] Hua K, Oishi K, Zhang J, Wakana S, Yoshioka T, Zhang W, Akhter KD, Li X, Huang H, Jiang H, van Zijl P, Mori S. Mapping of functional areas in the human cortex based on connectivity through association fibers. Cereb Cortex. 2009 Aug;19(8):1889-95. doi: 10.1093/cercor/bhn215. Epub 2008 Dec 9. PMID: 19068488; PMCID: PMC2705697.

[25] Powell HW, Parker GJ, Alexander DC, Symms MR, Boulby PA, Wheeler-Kingshott CA, Barker GJ, Noppeney U, Koepp MJ, Duncan JS. Hemispheric asymmetries in language-related pathways: a combined functional MRI and tractography study. Neuroimage. 2006 Aug 1;32(1):388-99. doi: 10.1016/j.neuroimage.2006.03.011. Epub 2006 May 2. PMID: 16632380.

[26] Saur D, Schelter B, Schnell S, Kratochvil D, Küpper H, Kellmeyer P, Kümmerer D, Klöppel S, Glauche V, Lange R, Mader W, Feess D, Timmer J, Weiller C. Combining functional and anatomical connectivity reveals brain networks for auditory language comprehension. Neuroimage. 2010 Feb 15;49(4):3187-97. doi: 10.1016/j.neuroimage.2009.11.009. Epub 2009 Nov 12. PMID: 19913624.

[27] Brauer J, Anwander A, Friederici AD. Neuroanatomical prerequisites for language functions in the maturing brain. Cereb Cortex. 2011 Feb;21(2):459-66. doi: 10.1093/cercor/bhq108. Epub 2010 Jun 21. PMID: 20566580.

[28] Yamada K. Diffusion tensor tractography should be used with caution. Proc Natl Acad Sci U S A. 2009 Feb 17;106(7):E14; author reply E15. doi: 10.1073/pnas.0812352106. Epub 2009 Jan 29. PMID: 19179404; PMCID: PMC2650181.

[29] Tremblay P, Dick AS. Broca and Wernicke are dead, or moving past the classic model of language neurobiology. Brain Lang. 2016 Nov;162:60-71. doi: 10.1016/j.bandl.2016.08.004. Epub 2016 Aug 30. PMID: 27584714.

[30] Axer H, von Keyserlingk AG, Berks G, von Keyserlingk DG. Supra- and infrasylvian conduction aphasia. Brain Lang. 2001 Mar;76(3):317-31. doi: 10.1006/brln.2000.2425. PMID: 11247647.

[31] Van Essen DC, Smith SM, Barch DM, Behrens TE, Yacoub E, Ugurbil K; WU-Minn HCP Consortium. The WU-Minn Human Connectome Project: an overview. Neuroimage. 2013 Oct 15;80:62-79. doi: 10.1016/j.neuroimage.2013.05.041. Epub 2013 May 16. PMID: 23684880; PMCID: PMC3724347.

[32] Sotiropoulos SN, Moeller S, Jbabdi S, Xu J, Andersson JL, Auerbach EJ, Yacoub E, Feinberg D, Setsompop K, Wald LL, Behrens TE, Ugurbil K, Lenglet C. Effects of image reconstruction on fiber orientation mapping from multichannel diffusion MRI: reducing the noise floor using SENSE. Magn Reson Med. 2013 Dec;70(6):1682-9. doi: 10.1002/mrm.24623. Epub 2013 Feb 7. PMID: 23401137; PMCID: PMC3657588.

[33] Glasser MF, Van Essen DC. Mapping human cortical areas in vivo based on myelin content as revealed by T1- and T2-weighted MRI. J Neurosci. 2011 Aug 10;31(32):11597-616. doi: 10.1523/JNEUROSCI.2180-11.2011. PMID: 21832190; PMCID: PMC3167149.

[34] Andersson JL, Skare S, Ashburner J. How to correct susceptibility distortions in spin-echo echo-planar images: application to diffusion tensor imaging. Neuroimage. 2003 Oct;20(2):870-88. doi: 10.1016/S1053-8119(03)00336-7. PMID: 14568458.

[35] Tournier JD, Calamante F, Gadian DG, Connelly A. Direct estimation of the fiber orientation density function from diffusion-weighted MRI data using spherical deconvolution. Neuroimage. 2004 Nov;23(3):1176-85. doi: 10.1016/j.neuroimage.2004.07.037. PMID: 15528117.

[36] Tournier JD, Calamante F, Connelly A. Robust determination of the fibre orientation distribution in diffusion MRI: non-negativity constrained super-resolved spherical deconvolution. Neuroimage. 2007 May 1;35(4):1459-72. doi: 10.1016/j.neuroimage.2007.02.016. Epub 2007 Feb 21. PMID: 17379540.

[37] Frank LR. Anisotropy in high angular resolution diffusion-weighted MRI. Magn Reson Med. 2001 Jun;45(6):935-9. doi: 10.1002/mrm.1125. PMID: 11378869.

[38] Tuch DS, Reese TG, Wiegell MR, Makris N, Belliveau JW, Wedeen VJ. High angular resolution diffusion imaging reveals intravoxel white matter fiber heterogeneity. Magn Reson Med. 2002 Oct;48(4):577-82. doi: 10.1002/mrm.10268. PMID: 12353272.

[39] Tournier JD, Yeh CH, Calamante F, Cho KH, Connelly A, Lin CP. Resolving crossing fibres using constrained spherical deconvolution: validation using diffusion-weighted imaging phantom data. Neuroimage. 2008 Aug 15;42(2):617-25. doi: 10.1016/j.neuroimage.2008.05.002. Epub 2008 May 9. PMID: 18583153.

[40] Tournier JD, Calamante F, Connelly A. Determination of the appropriate b value and number of gradient directions for high-angular-resolution diffusion-weighted imaging. NMR Biomed. 2013 Dec;26(12):1775-86. doi: 10.1002/nbm.3017. Epub 2013 Aug 29. PMID: 24038308.

[41] Tournier JD, Calamante F, Connelly A. MRtrix: diffusion tractography in crossing fiber regions. Int. J. Imaging Syst. Technol. 2012; 22: 53–66.

[42] Tournier JD, Mori S, Leemans A. Diffusion tensor imaging and beyond. Magn Reson Med. 2011 Jun;65(6):1532-56. doi: 10.1002/mrm.22924. Epub 2011 Apr 5. PMID: 21469191; PMCID: PMC3366862.

[43] Smith RE, Tournier JD, Calamante F, Connelly A. Anatomically-constrained tractography: improved diffusion MRI streamlines tractography through effective use of anatomical information. Neuroimage. 2012 Sep;62(3):1924-38. doi: 10.1016/j.neuroimage.2012.06.005. Epub 2012 Jun 13. PMID: 22705374.

[44] Smith RE, Tournier JD, Calamante F, Connelly A. The effects of SIFT on the reproducibility and biological accuracy of the structural connectome. Neuroimage. 2015 Jan 1;104:253-65. doi: 10.1016/j.neuroimage.2014.10.004. Epub 2014 Oct 12. PMID: 25312774.

[45] Tournier, J.-D.; Smith, R. E.; Raffelt, D.; Tabbara, R.; Dhollander, T.; Pietsch, M.; Christiaens, D.; Jeurissen, B.; Yeh, C.-H. & Connelly, A. MRtrix3: A fast, flexible and open software framework for medical image processing and visualisation. NeuroImage, 2019, 202, 116137

[46] Calamante F, Tournier JD, Jackson GD, Connelly A. Track-density imaging (TDI): super-resolution white matter imaging using whole-brain track-density mapping. Neuroimage. 2010 Dec;53(4):1233-43. doi:

10.1016/j.neuroimage.2010.07.024. Epub 2010 Jul 17. PMID: 20643215.

[47] Tianzi Jiang, Brainnetome: A new -ome to understand the brain and its disorders, NeuroImage, Volume 80, 2013, Pages 263-272, ISSN 1053-8119, https://doi.org/10.1016/j.neuroimage.2013.04.002.

[48] Fan L, Wang J, Zhang Y, Han W, Yu C, Jiang T. Connectivity-based parcellation of the human temporal pole using diffusion tensor imaging. Cereb Cortex. 2014 Dec;24(12):3365-78. doi: 10.1093/cercor/bht196. Epub 2013 Aug 7. PMID: 23926116.

[49] Fan L., Li H., Yu S., Jiang T. (2016) Human Brainnetome Atlas and Its Potential Applications in Brain-Inspired Computing. In: Amunts K., Grandinetti L., Lippert T., Petkov N. (eds) Brain-Inspired Computing. BrainComp 2015. Lecture Notes in Computer Science, vol 10087. Springer, Cham. https://doi.org/10.1007/978-3-319-50862-7_1

[50] Fischi B. (2012). FreeSurfer. NeuroImage, 62:774-781.

[51] Murray, E. A., Wise, S. P., & Graham, K. S. (2017). The evolution of memory systems: Ancestors, anatomy, and adaptations. Oxford University Press.

[52] IBM Corp., 2017. IBM SPSS Statistics for Windows, Armonk, NY: IBM Corp. Available at: https://hadoop.apache.org.

[53] Barbey AK, Koenigs M, Grafman J. Dorsolateral prefrontal contributions to human working memory. Cortex. 2013 May;49(5):1195-205. doi: 10.1016/j.cortex.2012.05.022. Epub 2012 Jun 16. PMID: 22789779; PMCID: PMC3495093.

[54] Na DG, Ryu JW, Byun HS, Choi DS, Lee EJ, Chung WI, Cho JM, Han BK. Functional MR imaging of working memory in the human brain. Korean J Radiol. 2000 Jan-Mar;1(1):19-24. doi: 10.3348/kjr.2000.1.1.19. PMID: 11752924; PMCID: PMC2718132.

[55] Baldo JV, Dronkers NF. The role of inferior parietal and inferior frontal cortex in working memory. Neuropsychology. 2006 Sep;20(5):529-38. doi: 10.1037/0894-4105.20.5.529. PMID: 16938015.

[56] Smith EE, Jonides J, Marshuetz C, Koeppe RA. Components of verbal working memory: evidence from neuroimaging. Proc Natl Acad Sci U S A. 1998 Feb 3;95(3):876-82. doi: 10.1073/pnas.95.3.876. PMID: 9448254; PMCID: PMC33811.

[57] Baddeley A. Working memory and language: an overview. J Commun Disord. 2003 May-Jun;36(3):189-208. doi: 10.1016/s0021-9924(03)00019-4. PMID: 12742667.

[58] Baddeley A. Working memory. Curr Biol. 2010 Feb 23;20(4):R136-40. doi: 10.1016/j.cub.2009.12.014. PMID: 20178752.

[59] Baddeley A, Hitch GJ. Working memory. In Bower GA, editor, Recent Advances in Learning and Motivation. Vol. 8. New York: Academic Press. 1974. p. 47-90

[60] Baddeley A. Working memory. Oxford psychology series. 1986 No.11. Clarendon Press/Oxford University Press.

[61] Binder JR. The Wernicke area: Modern evidence and a reinterpretation. Neurology. 2015 Dec 15;85(24):2170-5. doi: 10.1212/WNL.0000000000002219. Epub 2015 Nov 13. PMID: 26567270; PMCID: PMC4691684.

[62] Ding J, Martin RC, Hamilton AC, Schnur TT. Dissociation between frontal and temporal-parietal contributions to connected speech in acute stroke. Brain. 2020 Mar 1;143(3):862-876. doi: 10.1093/brain/awaa027. Erratum in: Brain. 2020 Jul 1;143(7):e63. PMID: 32155246; PMCID: PMC7089660.

[63] Ardila A, Bernal B, Rosselli M. Language and visual perception associations: meta-analytic connectivity modeling of Brodmann area 37. Behav Neurol. 2015;2015:565871. doi: 10.1155/2015/565871. Epub 2015 Jan 12. PMID: 25648869; PMCID: PMC4306224.

[64] Jouen AL, Ellmore TM, Madden CJ, Pallier C, Dominey PF, Ventre-Dominey J. Beyond the word and image: characteristics of a common meaning system for language and vision revealed by functional and structural imaging. Neuroimage. 2015 Feb 1;106:72-85. doi: 10.1016/j.neuroimage.2014.11.024. Epub 2014 Nov 15. PMID: 25463475.

[65] Brass M, Derrfuss J, Forstmann B, von Cramon DY. The role of the inferior frontal junction area in cognitive control. Trends Cogn Sci. 2005 Jul;9(7):314-6. doi: 10.1016/j.tics.2005.05.001. PMID: 15927520.

[66] Derrfuss J, Brass M, von Cramon DY. Cognitive control in the posterior frontolateral cortex: evidence from common activations in task coordination, interference control, and working memory. Neuroimage. 2004 Oct;23(2):604-12. doi: 10.1016/j.neuroimage.2004.06.007. PMID: 15488410.

[67] Derrfuss J, Brass M, Neumann J, von Cramon DY. Involvement of the inferior frontal junction in cognitive control: meta-analyses of switching and Stroop studies. Hum Brain Mapp. 2005 May;25(1):22-34. doi: 10.1002/hbm.20127. PMID: 15846824; PMCID: PMC6871679.

[68] Neumann J, Lohmann G, Derrfuss J, von Cramon DY. Meta-analysis of functional imaging data using replicator dynamics. Hum Brain Mapp. 2005 May;25(1):165-73. doi: 10.1002/hbm.20133. PMID: 15846812; PMCID: PMC6871715.

[69] Townsend J, Bookheimer SY, Foland-Ross LC, Sugar CA, Altshuler LL. fMRI abnormalities in dorsolateral prefrontal cortex during a working memory task in manic, euthymic and depressed bipolar subjects. Psychiatry Res. 2010 Apr 30;182(1):22-9. doi: 10.1016/j.pscychresns.2009.11.010. Epub 2010 Mar 15. PMID: 20227857; PMCID: PMC2918407.

[70] Snow PJ. The Structural and Functional Organization of Cognition. Front Hum Neurosci. 2016 Oct 17;10:501. doi: 10.3389/fnhum.2016.00501. PMID: 27799901; PMCID: PMC5065967.

[71] Kinoshita M, Shinohara H, Hori O, Ozaki N, Ueda F, Nakada M, Hamada J, Hayashi Y. Association fibers connecting the Broca center and the lateral superior frontal gyrus: a microsurgical and tractographic anatomy. J Neurosurg. 2012 Feb;116(2):323-30. doi: 10.3171/2011.10.JNS11434. Epub 2011 Nov 11. PMID: 22077454.

**Figure legends**

**Figure 1.** Summary of the: **[A]** three-segment model of SLF connectivity and the **[B]** two-segment model of AF connectivity. AG, angular gyrus; IFG, inferior frontal gyrus; MFG, middle frontal gyrus; MITG, middle and inferior temporal gyri; SFG, superior frontal gyrus; Pcun, precuneus; SMG, supramarginal gyrus; SMTG, superior and middle temporal gyrus

**Figure 2. [A]** The workflow of the whole data processing steps. **[B]** Connectome process based on the parcellated Brainnetome atlas node for regions of interest.

**Figure 3.** Reconstruct fiber bundles using the high angular resolution diffusion-weighted imaging strategy **[A]** PFC-IPC fiber tracks **[B]** PFC-TC fiber tracts. IPC, inferior parietal cortex; PFC, prefrontal cortex; TC, temporal cortex

**Figure** 1 **[A]** Average track-density matrix of the acquired fiber tracts from the PFC-IPC. **[B]** Box-and-whisker plot showing the IFJ-IPC connectivity specifically for the regions indicated in panel A. **[C]** Box-and-whisker plot showing the A44v connectivity specifically for the regions indicated in the panel A. **[D]** Comparison of one-sample t-values of the connectivity between the selected subregions within the PFC-IPC. In the matrix, the connectivities with p-values greater than 0.05 are indicated as zero. **[E]** Three-dimensional visualization of track density in PFC-IPC connectivity (Supplementary Table 2). A8m, medial area 8; A8dl, dorsolateral area 8; A8vl, ventrolateral area 8; A9l, lateral area 9; A9m, medial area 9; A10m, medial area 10; A9/46d, dorsal area 9/46; A10l, lateral area 10; A11l, lateral area 11; A11m, medial area 11; A12/47l, lateral area 12/47; A12/47o, opercular area 12/47; A13, area 13; A14m, medial area 14; A39c, caudal area 39; A39rd, rostrodorsal area 39; A39rv, rostroventral area 39; A40c, caudal area 40; A40rd, rostrodorsal area 40; A40rv, rostroventral area 40; A44d, dorsal area 44; A44op, opercular area 44; A44v, ventral area 44; A45c, caudal area 45; A45r, rostral area 45; A46, area 46; IFJ, inferior frontal junction; IFS, inferior frontal sulcus

**Figure 5. [A]** Average track-density matrix of the acquired fiber tracts from the PFC-TC. **[B]** Box-and-whisker plot showing the IFJ-TC connectivity specifically for the regions indicated in panel A. **[C]** Box-and-whisker plot showing the A44v-TC connectivity specifically for the regions indicated in panel A. **[D]** Comparison of one sample t-values of the connectivity between the selected subregions within the PFC-TC. In the matrix, the connectivities that have p-values greater than 0.05 are indicated as zero. **[E]** Three-dimensional visualization of track density in PFC-TC connectivity (Supplementary Table 4). A8m, medial area 8; A8dl, dorsolateral area 8; A8vl, ventrolateral area 8; A9l, lateral area 9; A9m, medial area 9; A9/46d, dorsal area 9/46; A10m, medial area 10; A10l, lateral area 10; A11l, lateral area 11; A11m, medial area 11; A12/47l, lateral area 12/47; A12/47o, opercular area 12/47; A13, area 13; A14m, medial area 14; A21c, caudal area 21; A21r, rostral area 21; A22c, caudal area 22; A22r, rostral area 22; A37dl, dorsolateral area 37; A38l, lateral area 38; A38m, medial area 38; A41/42, area 41/42; A44d, dorsal area 44; A44op, opercular area 44; A44v, ventral area 44; A45c, caudal area 45; A45r, rostral area 45; A46, area 46; aSTS, anterior superior temporal sulcus; IFJ, inferior frontal junction; IFS, inferior frontal sulcus; TE, TE 1.0 and TE 1.2

**Figure 6.** Mapping the endpoints of the streamline to the IPC **[A]** IFJ-IPC streamlines and **[B]** A44v-IPC streamlines. A8m, medial area 8; A8dl, dorsolateral area 8; A8vl, ventrolateral area 8; A9l, lateral area 9; A9m, medial area 9; A9/46d, dorsal area 9/46; A10l, lateral area 10; A10m, medial area 10; A11l, lateral area 11; A11m, medial area 11; A12/47l, lateral area 12/47; A12/47o, opercular area 12/47; A13, area 13; A14m, medial area 14; A39c, caudal area 39; A39rd, rostrodorsal area 39; A39rv, rostroventral area 39; A40c, caudal area 40; A40rd, rostrodorsal area 40; A40rv, rostroventral area; 40A46, area 46; A44d, dorsal area 44; A44op, opercular area 44; A44v, ventral area 44; A45c, caudal area 45; A45r, rostral area 45; IFJ, inferior frontal junction; IFS, inferior frontal sulcus

**Figure 7.** Mapping the endpoints of the streamlines to the TC: **[A]** IFJ-TC streamlines; and **[B]** A44v-TC streamlines. A8dl, dorsolateral area 8; A8m, medial area 8; A8vl, ventrolateral area 8; A9l, lateral area 9; A9m, medial area 9; A9/46d, dorsal area 9/46; A10l, lateral area 10; A10m, medial area 10; A11l, lateral area 11; A11m, medial area 11; A12/47l, lateral area 12/47; A12/47o, opercular area 12/47; A14m, medial area 14; A13, area 13; A21c, caudal area 21; A21r, rostral area 21; A22c, caudal area 22; A22r, rostral area 22; A37dl, dorsolateral area 37; A38l, lateral area 38; A38m, medial area 38; A41/42, area 41/42; A44d, dorsal area 44; A44op, opercular area 44; A44v, ventral area 44; A45c, caudal area 45; A45r, rostral area 45; A46, area 46; aSTS, anterior superior temporal

sulcus; IFJ, inferior frontal junction; IFS, inferior frontal sulcus; TE, TE 1.0 and TE 1.2

**Figure 8.** [A] Average track density matrix of the acquired short fiber tracts from the PFC. [B] Box-and-whisker plot showing the PFC-IFJ connectivity specifically for the regions indicated in panel A. [C] Box-and-whisker plot showing the PFC-A44v connectivity specifically for the regions indicated in panel A. [D] Comparison of one-sample t-values of the connectivity between the short fibers within the PFC. In the matrix, the connectivities that have p-values greater than 0.05 were indicated as zero. [D] Three-dimensional visualization of track density within PFC-IFJ focused short fiber connectivity (Supplementary Table 5). A8dl, dorsolateral area 8; A8m, medial area 8; A9l, lateral area 9; A9m, medial area 9; A9/46d, dorsal area 9/46; A10m, medial area 10; A8vl, ventrolateral area 8; A10l, lateral area 10; A11l, lateral area 11; A11m, medial area 11; A12/47l, lateral area 12/47; A12/47o, opercular area 12/47; A13, area 13; A14m, medial area 14; A44d, dorsal area 44; A45c, caudal area 45; A45r, rostral area 45; A44op, opercular area 44; A44v, ventral area 44; A46, area 46; IFJ, inferior frontal junction; IFS, inferior frontal sulcus

**Figure 9.** Summary of the results: [A] PFC-IPC connectivity; [B] PFC-TC connectivity; [C] A block diagram of the structural connectivity of the dorsal language pathway. A44v, ventral area of BA44; AF, arcuate fasciculus; AG, angular gyrus; DLPFC, dorsolateral prefrontal cortex; FG, fusiform gyrus; IFJ, inferior frontal junction; IPC, inferior parietal cortex; PFC, prefrontal cortex; SLF, superior longitudinal fasciculus; SMG, supramarginal gyrus; TC, temporal cortex

**Table 1.** Contemporary and sometimes contentious models of the superior longitudinal fasciculus (SLF) and arcuate fasciculus (AF) connectivity.

**Table 2.** Guideline for reconstructing fiber tracts.

**Table 3.** Guideline for reconstructing fiber tracts originating from the major prefrontal cortex (PFC) subregions.

**Figures & Tables**

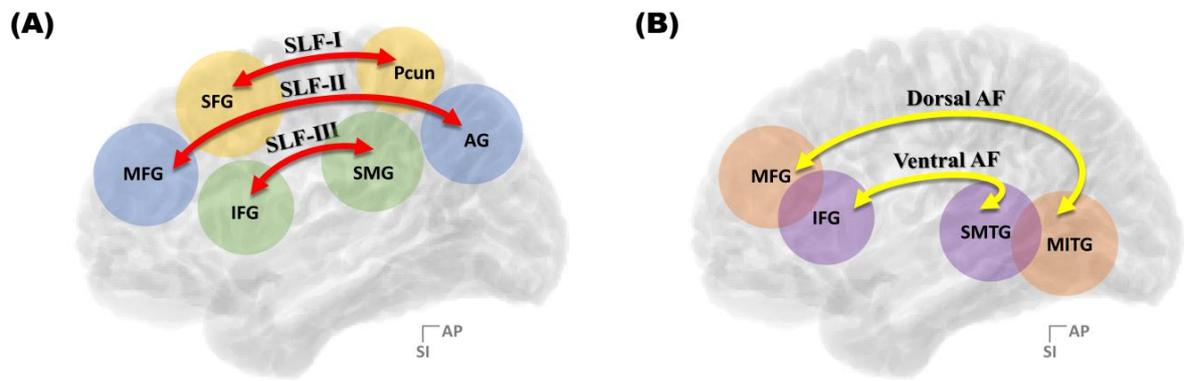

**Figure 1.**

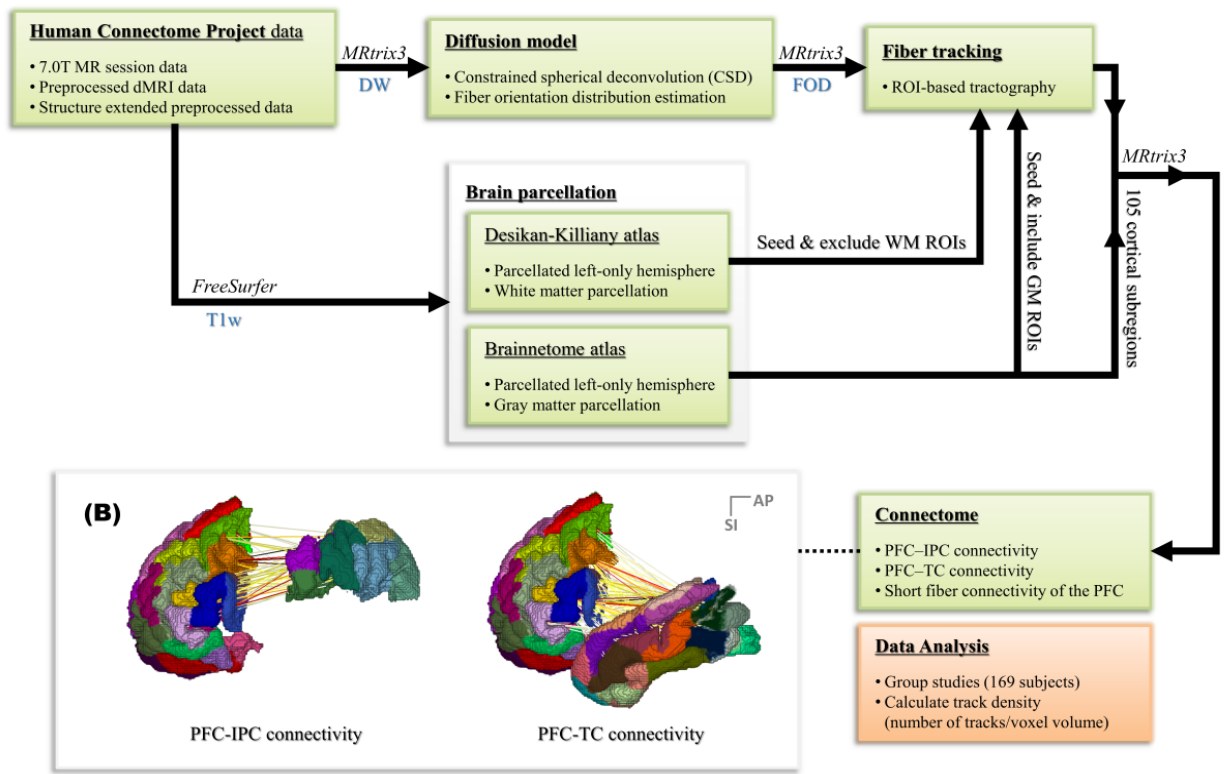

**Figure 2.**

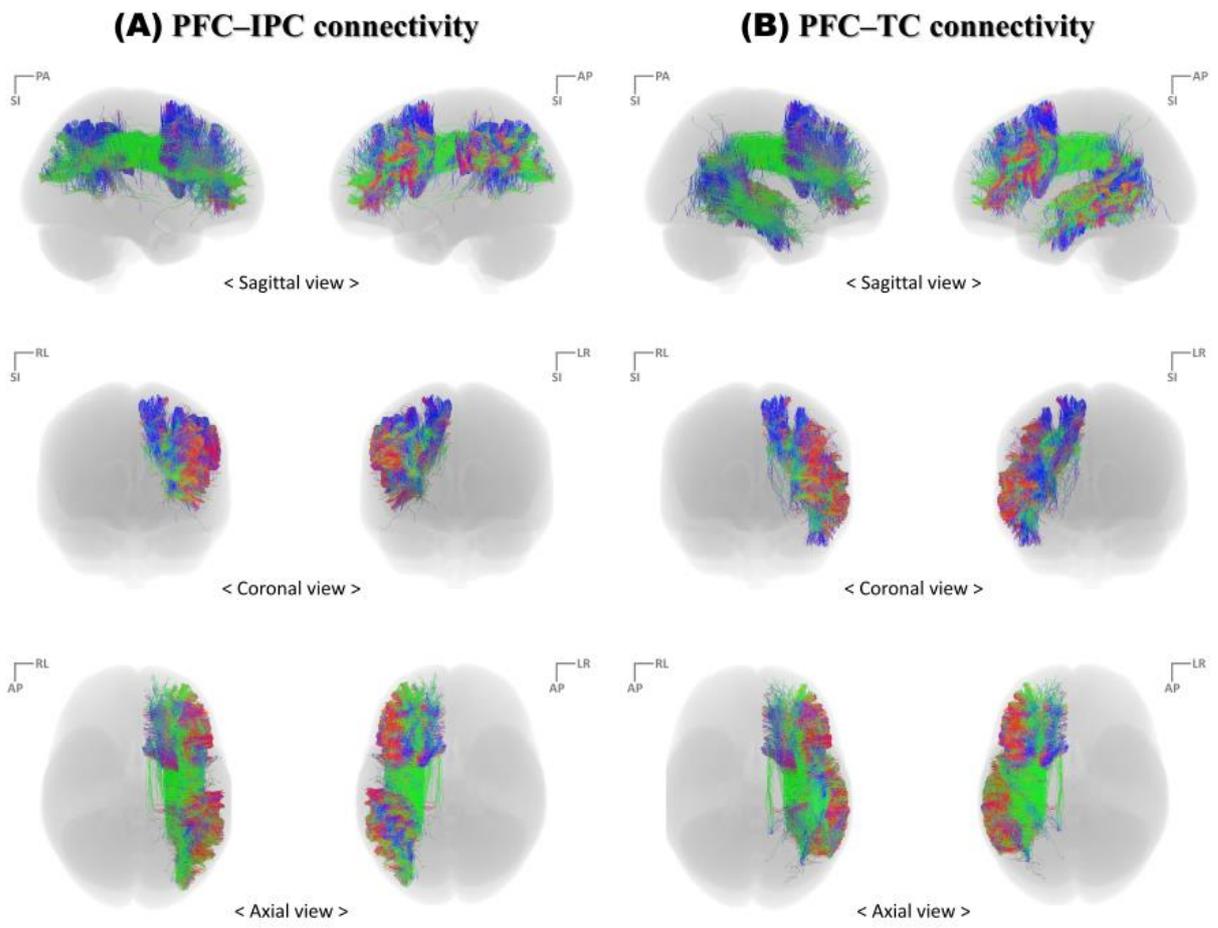

**Figure 3.**

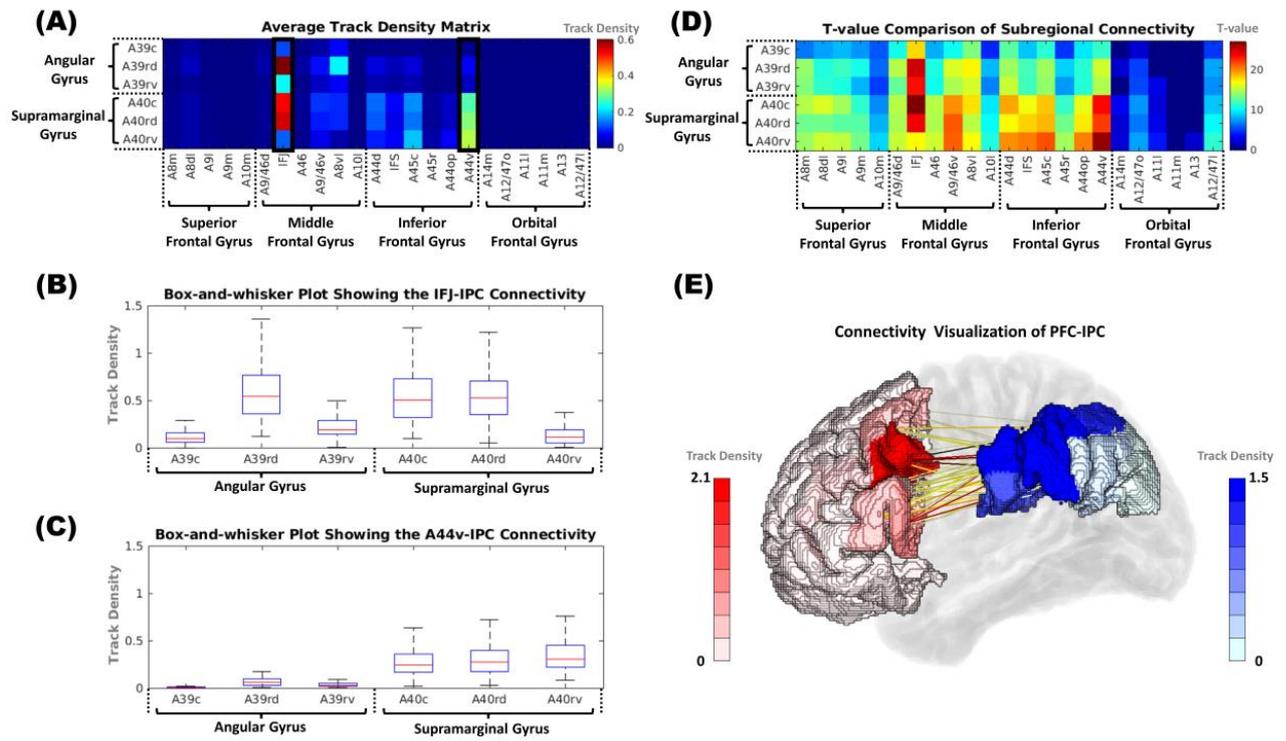

**Figure 4.**

**Figure 5.**

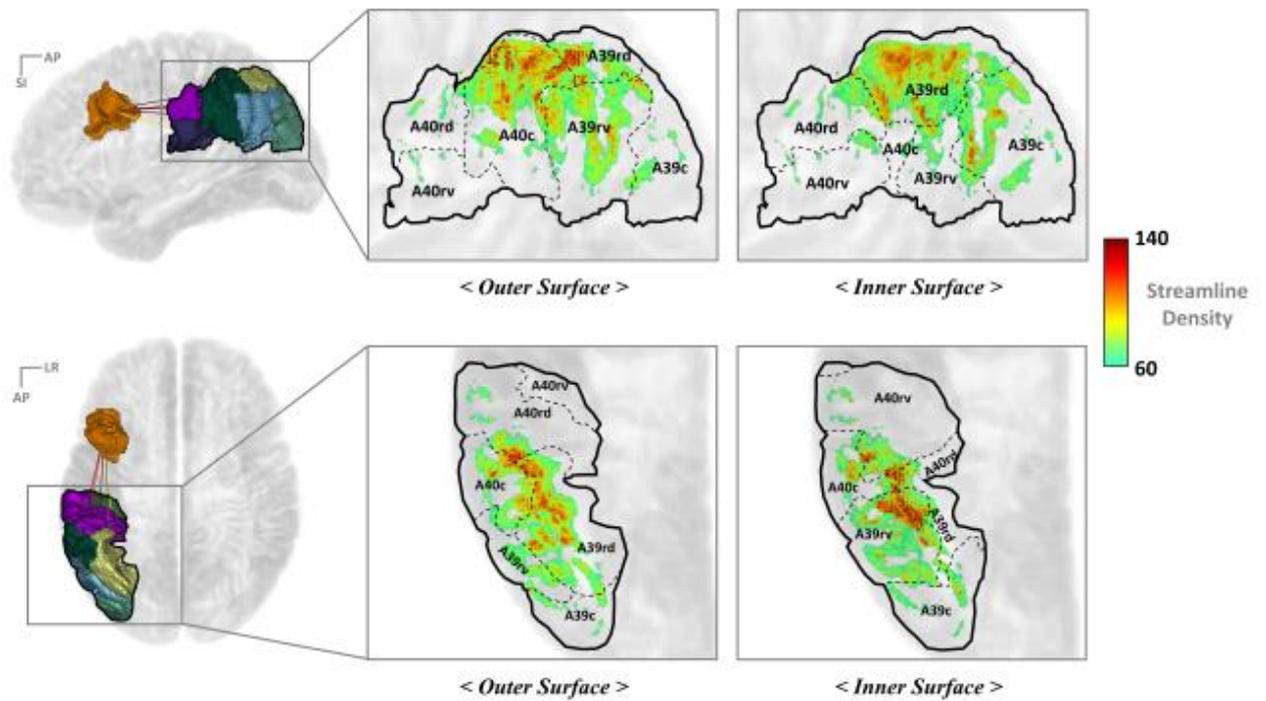
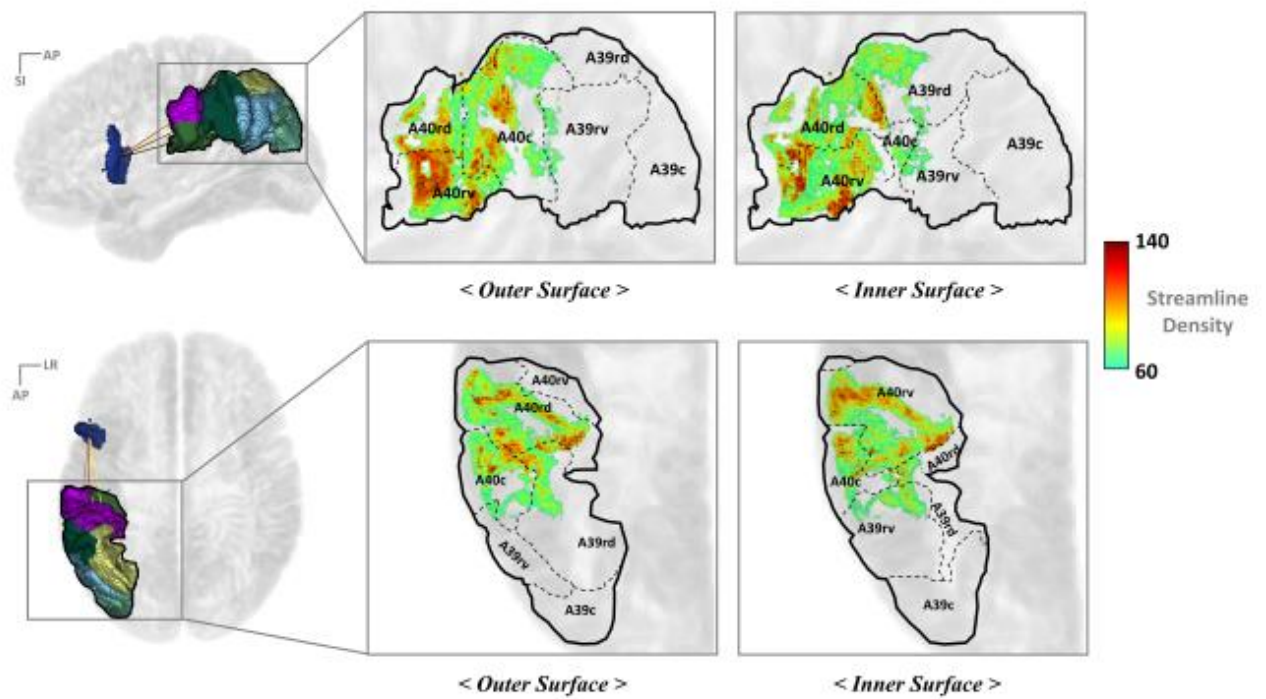

**Figure 6.**

**(A)** Mapping the endpoints of the IFJ-TC streamlines to the TC

**(B)** Mapping the endpoints of the A44v-TC streamlines to the TC

**Figure 7.**

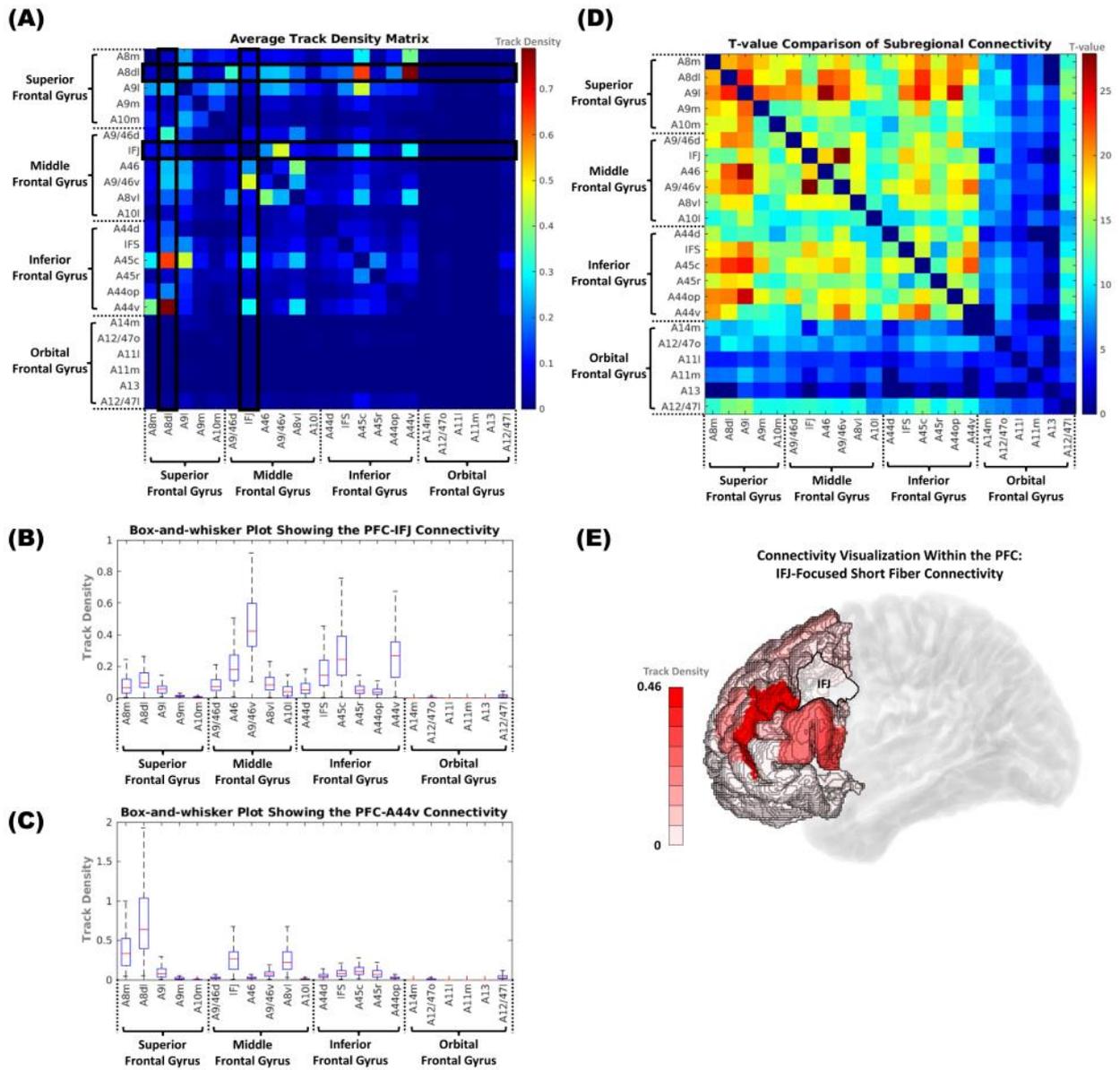

**Figure 8.**

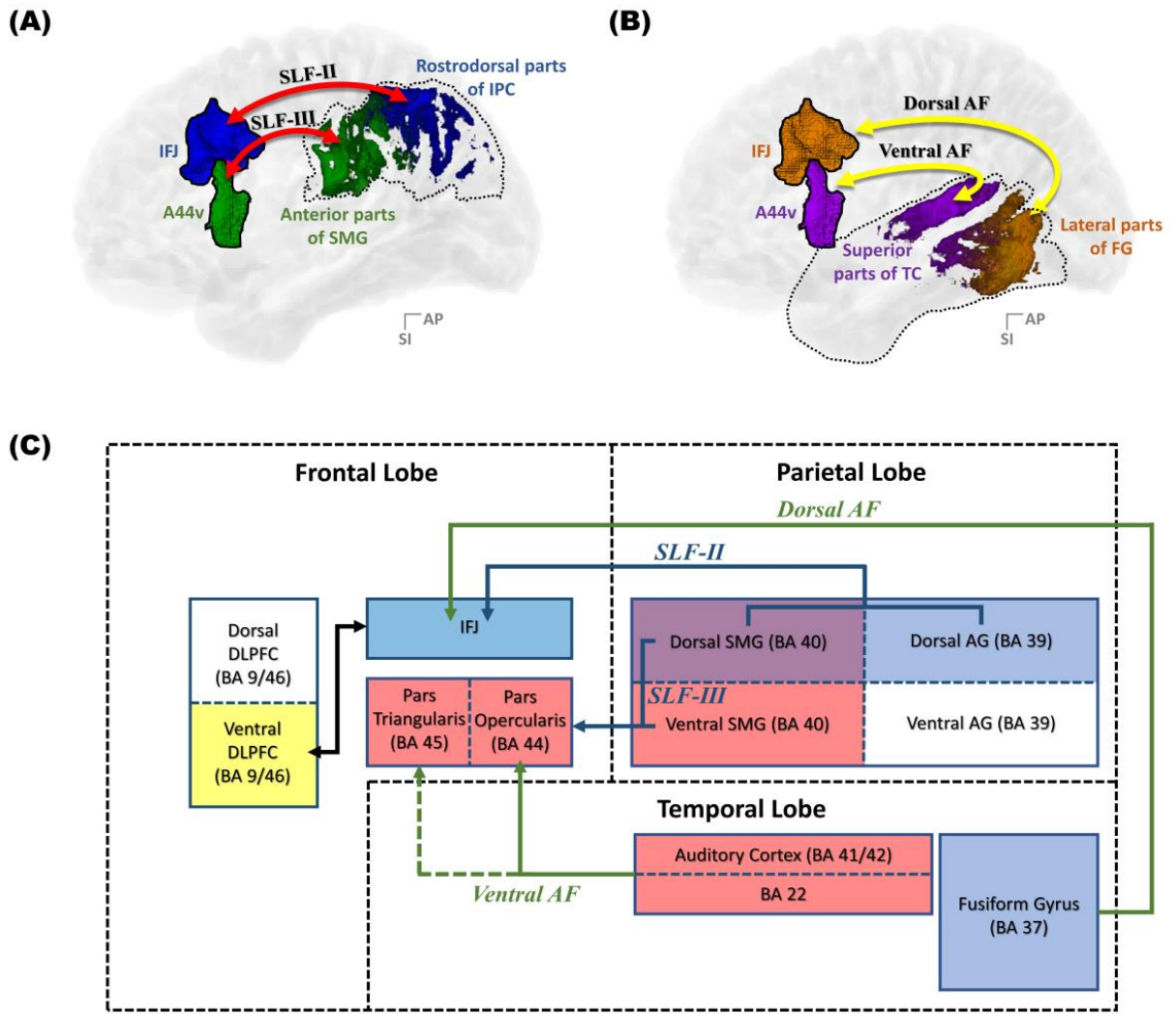

**Figure 9.**

**Table. 1. Contemporary and sometimes contentious models regarding the SLF and AF connectivity.**

| Study Groups | Fiber Tracts | Interconnected Regions | |
| --- | --- | --- | --- |
| | | **Originated Regions** | **Terminated Regions** |
| Catani et al. (2005) | SLF and AF | Direct Temporo-frontal Pathway and Indirect Temporal-parietal-frontal Pathway | |
| Makris et al. (2005) | SLF-I | Superior and Medial Parietal Cortex | Dorsal Parts of the Frontal Region |
| | SLF-II | Posterior-inferior Parietal Region | Prefrontal Region |
| | SLF-III | Supramarginal Gyrus | Brodmann's Area of 6, 44, and 46 |
| | AF | Caudal Superior Temporal Region | Brodmann's Area of 8 and 46 |
| Thiebaut de Schotten et al. (2011) | SLF-I | Precuneus and the Superior Parietal Lobe | Superior Frontal and Anterior Cingulate Gyri |
| | SLF-II | Anterior Intermediate Parietal Sulcus and the Angular Gyrus | Posterior Parts of the Superior and Middle Frontal Gyri |
| | SLF-III | Temporo-parietal Junction | Inferior Frontal Gyrus |
| Kamali et al. (2014) | SLF-I | Superior and Medial Parietal Cortex | Dorsal and Medial Cortex of the Frontal Lobe |
| | SLF-II | Posteriolateral Parts of the Parietal Cortex | Dorsolateral Prefrontal Cortex |
| | SLF-III | Rostral Portion of the Inferior Paretal Lobe | Ventral Premotor and Prefrontal Cortex |
| | AF | Superior Temporal Gyrus | Dorsolateral Prefrontal Cortex |
| Wang et al. (2015) and Yagmurlu et al. (2016) | SLF-I | Precuneus | Anterior Cingulate Cortex and Posterior Parts of the Superior Frontal Gyrus |
| | SLF-II | Angular Gyrus | Middle Frontal Gyrus |
| | SLF-III | Supramarginal Gyrus | Pars Opercularis |
| Glasser and Rilling (2008) and Yagmurlu et al. (2016) | dorsal-AF | Posterior Parts of the Middle and Inferior Temporal Gyrus | Ventral Premotor Cortex, Pars Opercularis, Posterior Middle Frontal Gyrus |
| | ventral-AF | Middle and Posterior Parts of the Superior Temporal Gyrus | Ventral Premotor Cortex, Pars Opercularis and Triangularis |

**Table 1.**

**Table. 2. Guideline for reconstructing main fiber tracts**

| Fiber tract | Streamline Seeding Region (White Matter) | Included Region (Gray Matter) | Excluded Regions (grey matter) | Number of Tracks Per Subject | Reference Figure |
|---|---|---|---|---|---|
| PFC-IPC connectivity (SLF-II and SLF-III) | Inferior Parietal Gyrus | Prefrontal Cortex | Temporal Cortex, Insular Cortex, Limbic Cortex, Occipital Cortex, and Subcortex | 10000 | Fig 3(A) |
| PFC-TC connectivity (AF) | Temporal Gyrus | Prefrontal Cortex | Parietal Cortex, Insular Cortex, Limbic Cortex, Occipital Cortex, and Subcortex | 10000 | Fig 3(B) |
| Short fiber connectivity within the PFC | Frontal Lobe | Prefrontal Cortex | Temporal Cortex, Parietal Cortex, Insular Cortex, Limbic Cortex, Occipital Cortex, and Subcortex | 10000 | Fig 3(C) |

**Table 2.**

**Table. 3. Guideline for reconstructing fiber tracts originated from the major PFC subregions**

| Fiber tract | Streamline Seeding Region (Gray Matter) | Included Region (Gray Matter) | Excluded Regions (grey matter) | Number of Tracks Per Subject |
|---|---|---|---|---|
| **IFJ-IPC connectivity** | Inferior Frontal Junction | Inferior Parietal Cortex | Temporal Cortex, Insular Cortex, Limbic Cortex, Occipital Cortex, and Subcortex | 10000 |
| **A44v-IPC connectivity** | A44v (ventral area of BA44) | Inferior Parietal Cortex | Temporal Cortex, Insular Cortex, Limbic Cortex, Occipital Cortex, and Subcortex | 10000 |
| **IFJ-TC connectivity** | Inferior Frontal Junction | Temporal Cortex | Temporal Cortex, Parietal Cortex, Insular Cortex, Limbic Cortex, Occipital Cortex, and Subcortex | 10000 |
| **A44v-TC connectivity** | A44v (ventral area of BA44) | Temporal Cortex | Temporal Cortex, Parietal Cortex, Insular Cortex, Limbic Cortex, Occipital Cortex, and Subcortex | 10000 |

**Table 3.**

**Supplementary Figures and Tables**

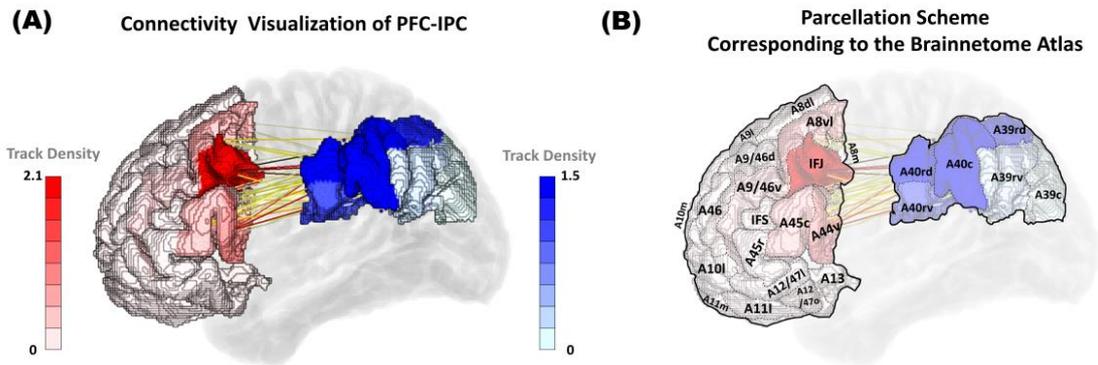

**Supplementary Figure 2. [A]** Connectivity visualization of PFC-IPC **[B]** Parcellation scheme corresponding to the Brainnetome atlas.

(**Abbreviation**: A8m(medial area 8), A8dl(dorsolateral area 8), A9l(lateral area 9), A9m(medial area 9), A10m(medial area 10), A9/46d(dorsal area 9/46), IFJ(inferior frontal junction), A46(area 46), A8vl(ventrolateral area 8), A10l(lateral area 10), A44d(dorsal area 44), IFS(inferior frontal sulcus), A45c(caudal area 45), A45r(rostral area 45), A44op(opercular area 44), A44v(ventral area 44), A14m(medial area 14), A12/47o(opercular area 12/47), A11l(lateral area 11), A11m(medial area 11), A13(area 13), A12/47l(lateral area 12/47), A39c(caudal area 39), A39rd(rostrodorsal area 39), A40rd(rostrodorsal area 40), A40c(caudal area 40), A39rv(rostroventral area 39), A40rv(rostroventral area 40))

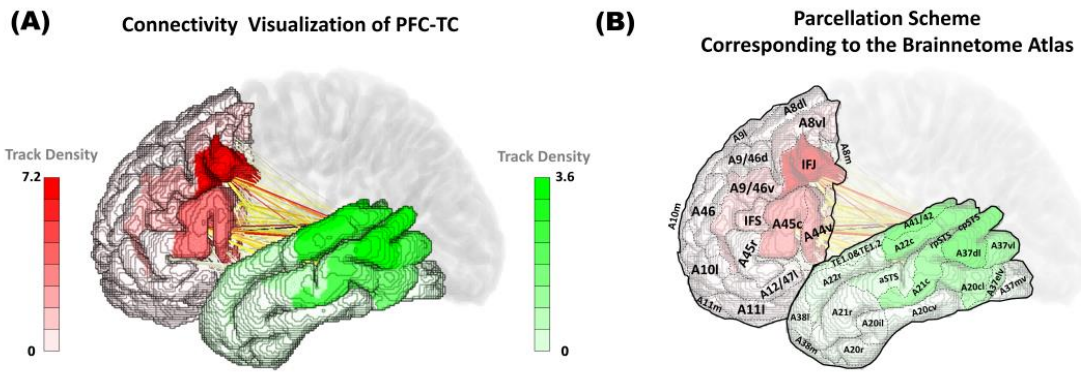

**Supplementary Figure 2. [A]** Connectivity visualization of PFC-TC **[B]** Parcellation scheme corresponding to the Brainnetome atlas.

(**Abbreviation**: A8m(medial area 8), A8dl(dorsolateral area 8), A9l(lateral area 9), A9m(medial area 9), A10m(medial area 10), A9/46d(dorsal area 9/46), IFJ(inferior frontal junction), A46(area 46), A8vl(ventrolateral area 8), A10l(lateral area 10), A44d(dorsal area 44), IFS(inferior frontal sulcus), A45c(caudal area 45), A45r(rostral area 45), A44op(opercular area 44), A44v(ventral area 44), A14m(medial area 14), A12/47o(opercular area 12/47), A11l(lateral area 11), A11m(medial area 11), A13(area 13), A12/47l(lateral area 12/47), A38m(medial area 38), A41/42(area 41/42), TE(TE1.0 and TE 1.2), A22c(caudal area 22), A38l(lateral area 38), A22r(rostral area 22), A21c(caudal area 21), A21r(rostral area 21), A37dl(dorsolateral area 37), aSTS(anterior superior temporal sulcus))

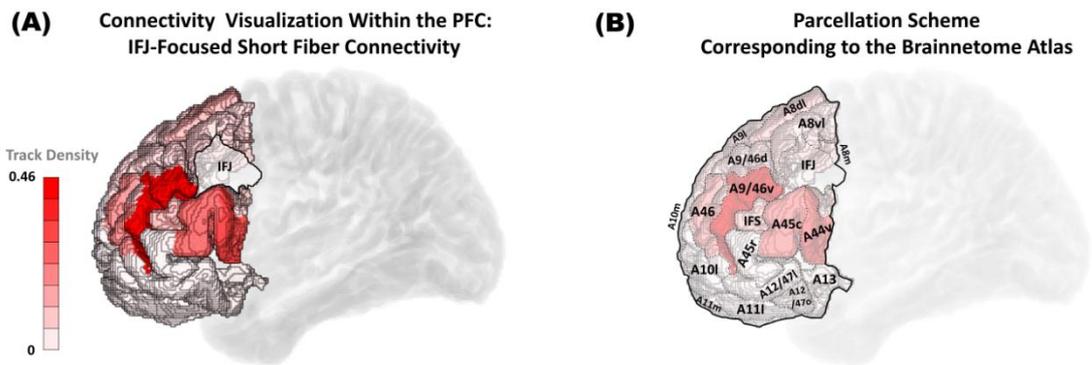

**Supplementary Figure 3. [A]** Connectivity visualization within the PFC: IFJ focused short fiber connectivity **[B]** Parcellation scheme corresponding to the Brainnetome atlas.

(**Abbreviation**: A8m(medial area 8), A8dl(dorsolateral area 8), A9l(lateral area 9), A9m(medial area 9), A10m(medial area 10), A9/46d(dorsal area 9/46), IFJ(inferior frontal junction), A46(area 46), A8vl(ventrolateral area 8), A10l(lateral area 10), A44d(dorsal area 44), IFS(inferior frontal sulcus), A45c(caudal area 45), A45r(rostral area 45), A44op(opercular area 44), A44v(ventral area 44), A14m(medial area 14), A12/47o(opercular area 12/47), A11l(lateral area 11), A11m(medial area 11), A13(area 13), A12/47l(lateral area 12/47))

**Supplementary Table 1.** Track density percentile for the PFC subregions regarding the PFC-IPC connectivity

| PFC Subregions | Total Track Density for the Reached Streamlines | Percentage (%) | Percentile (%ile) |
|---|---|---|---|
| A8m | 0.03473607 | 0.57 | 45.4 |
| A8dl | 0.127331018 | 2.11 | 63.6 |
| A9l | 0.022523596 | 0.37 | 40.9 |
| A9m | 0.005810138 | 0.10 | 27.2 |
| A10m | 0.002268416 | 0.04 | 22.7 |
| A9/46d | 0.07411429 | 1.23 | 59 |
| IFJ | 2.162089647 | 35.77 | 100 |
| A46 | 0.061958413 | 1.03 | 54.5 |
| A9/46v | 0.39671128 | 6.56 | 77.2 |
| A8vl | 0.587589996 | 9.72 | 90.9 |
| A10l | 0.015886064 | 0.26 | 36.3 |
| A44d | 0.444044835 | 7.35 | 81.8 |
| IFS | 0.261219566 | 4.32 | 72.7 |
| A45c | 0.580249838 | 9.60 | 86.3 |
| A45r | 0.056833528 | 0.94 | 50 |
| A44op | 0.16354551 | 2.71 | 68.1 |
| A44v | 1.031314328 | 17.06 | 95.4 |
| A14m | 0.000104186 | 0.00 | 9 |
| A12/47o | 0.002193343 | 0.04 | 18.1 |
| A11l | 0.000233674 | 0.00 | 13.6 |
| A11m | 2.57325E-05 | 0.00 | 0 |
| A13 | 5.12121E-05 | 0.00 | 4.5 |
| A12/47l | 0.013711607 | 0.23 | 31.8 |

**Supplementary Table 1.** Track density percentile for the PFC subregions regarding the PFC-IPC connectivity.

**Supplementary Table 2. One sample t-test results of the connectivity between the selected subregions within the PFC-IPC**

| Track Density | T | df | Sig. (2-tailed) | Mean Difference | 95% Confidence Interval of the Difference Lower | 95% Confidence Interval of the Difference Upper | Track Density | T | df | Sig. (2-tailed) | Mean Difference | 95% Confidence Interval of the Difference Lower | 95% Confidence Interval of the Difference Upper |
|---|---|---|---|---|---|---|---|---|---|---|---|---|---|
| A8m - A39c connectivity | 6.528 | 168 | 0.00000 | 0.00336 | 0.00230 | 0.00440 | A44d - A40rd connectivity | 16.913 | 168 | 0.00000 | 0.12757 | 0.11270 | 0.14250 |
| A8m - A39rd connectivity | 13.988 | 168 | 0.00000 | 0.00496 | 0.00430 | 0.00570 | A44d - A40c connectivity | 19.212 | 168 | 0.00000 | 0.13623 | 0.12220 | 0.15020 |
| A8m - A40rd connectivity | 13.08 | 168 | 0.00000 | 0.00799 | 0.00680 | 0.00920 | A44d - A39rv connectivity | 15.381 | 168 | 0.00000 | 0.02810 | 0.02450 | 0.03170 |
| A8m - A40c connectivity | 14.839 | 168 | 0.00000 | 0.00797 | 0.00690 | 0.00900 | A44d - A40rv connectivity | 19.807 | 168 | 0.00000 | 0.09061 | 0.08160 | 0.09960 |
| A8m - A39rv connectivity | 13.623 | 168 | 0.00000 | 0.00218 | 0.00190 | 0.00250 | IFJ - A39c connectivity | 8.643 | 168 | 0.00000 | 0.00515 | 0.00400 | 0.00630 |
| A8m - A40rv connectivity | 14.232 | 168 | 0.00000 | 0.00827 | 0.00710 | 0.00940 | IFJ - A39rd connectivity | 11.918 | 168 | 0.00000 | 0.03144 | 0.02620 | 0.03660 |
| A8dl - A39c connectivity | 7.667 | 168 | 0.00000 | 0.01764 | 0.01310 | 0.02220 | IFJ - A40rd connectivity | 18.025 | 168 | 0.00000 | 0.06060 | 0.05400 | 0.06720 |
| A8dl - A39rd connectivity | 10.94 | 168 | 0.00000 | 0.03327 | 0.02730 | 0.03930 | IFJ - A40c connectivity | 16.524 | 168 | 0.00000 | 0.07411 | 0.06530 | 0.08300 |
| A8dl - A40rd connectivity | 13.102 | 168 | 0.00000 | 0.02235 | 0.01900 | 0.02570 | IFJ - A39rv connectivity | 11.744 | 168 | 0.00000 | 0.01724 | 0.01430 | 0.02010 |
| A8dl - A40c connectivity | 15.858 | 168 | 0.00000 | 0.02209 | 0.01930 | 0.02480 | IFJ - A40rv connectivity | 20.28 | 168 | 0.00000 | 0.07268 | 0.06560 | 0.07980 |
| A8dl - A39rv connectivity | 11.378 | 168 | 0.00000 | 0.01042 | 0.00860 | 0.01220 | A45c - A39c connectivity | 11.216 | 168 | 0.00000 | 0.00770 | 0.00630 | 0.00910 |
| A8dl - A40rv connectivity | 15.497 | 168 | 0.00000 | 0.02155 | 0.01880 | 0.02430 | A45c - A39rd connectivity | 14.306 | 168 | 0.00000 | 0.04959 | 0.04270 | 0.05640 |
| A9l - A39c connectivity | 8.337 | 168 | 0.00000 | 0.00156 | 0.00120 | 0.00190 | A45c - A40rd connectivity | 18.754 | 168 | 0.00000 | 0.14192 | 0.12700 | 0.15690 |
| A9l - A39rd connectivity | 11.702 | 168 | 0.00000 | 0.00394 | 0.00330 | 0.00460 | A45c - A40c connectivity | 18.352 | 168 | 0.00000 | 0.15998 | 0.14280 | 0.17720 |
| A9l - A40rd connectivity | 12.492 | 168 | 0.00000 | 0.00391 | 0.00330 | 0.00450 | A45c - A39rv connectivity | 14.018 | 168 | 0.00000 | 0.02818 | 0.02420 | 0.03220 |
| A9l - A40c connectivity | 14.387 | 168 | 0.00000 | 0.00497 | 0.00430 | 0.00560 | A45c - A40rv connectivity | 21.454 | 168 | 0.00000 | 0.19288 | 0.17510 | 0.21060 |
| A9l - A39rv connectivity | 10.818 | 168 | 0.00000 | 0.00159 | 0.00130 | 0.00190 | A45r - A39c connectivity | 4.677 | 168 | 0.00000 | 0.00086 | 0.00050 | 0.00120 |
| A9l - A40rv connectivity | 15.108 | 168 | 0.00000 | 0.00656 | 0.00570 | 0.00740 | A45r - A39rd connectivity | 8.91 | 168 | 0.00000 | 0.00417 | 0.00320 | 0.00510 |
| A9m - A39c connectivity | 7.35 | 168 | 0.00000 | 0.00039 | 0.00030 | 0.00050 | A45r - A40rd connectivity | 13.075 | 168 | 0.00000 | 0.01108 | 0.00940 | 0.01280 |
| A9m - A39rd connectivity | 10.751 | 168 | 0.00000 | 0.00079 | 0.00060 | 0.00090 | A45r - A40c connectivity | 11.544 | 168 | 0.00000 | 0.01486 | 0.01230 | 0.01740 |
| A9m - A40rd connectivity | 12.371 | 168 | 0.00000 | 0.00136 | 0.00110 | 0.00160 | A45r - A39rv connectivity | 7.432 | 168 | 0.00000 | 0.00273 | 0.00200 | 0.00340 |
| A9m - A40c connectivity | 11.319 | 168 | 0.00000 | 0.00126 | 0.00100 | 0.00150 | A45r - A40rv connectivity | 14.45 | 168 | 0.00000 | 0.02314 | 0.02000 | 0.02630 |
| A9m - A39rv connectivity | 8.232 | 168 | 0.00000 | 0.00032 | 0.00020 | 0.00040 | A44op - A39c connectivity | 9.564 | 168 | 0.00000 | 0.00161 | 0.00130 | 0.00190 |
| A9m - A40rv connectivity | 13.141 | 168 | 0.00000 | 0.00170 | 0.00140 | 0.00200 | A44op - A39rd connectivity | 11.125 | 168 | 0.00000 | 0.00929 | 0.00760 | 0.01090 |
| A10m - A39c connectivity | 5.669 | 168 | 0.00000 | 0.00017 | 0.00010 | 0.00020 | A44op - A40rd connectivity | 17.377 | 168 | 0.00000 | 0.03699 | 0.03280 | 0.04120 |
| A10m - A39rd connectivity | 6.066 | 168 | 0.00000 | 0.00041 | 0.00030 | 0.00050 | A44op - A40c connectivity | 16.312 | 168 | 0.00000 | 0.03833 | 0.03370 | 0.04300 |
| A10m - A40rd connectivity | 8.044 | 168 | 0.00000 | 0.00044 | 0.00030 | 0.00050 | A44op - A39rv connectivity | 11.012 | 168 | 0.00000 | 0.00524 | 0.00430 | 0.00620 |
| A10m - A40c connectivity | 7.061 | 168 | 0.00000 | 0.00052 | 0.00040 | 0.00070 | A44op - A40rv connectivity | 20.177 | 168 | 0.00000 | 0.07208 | 0.06500 | 0.07910 |
| A10m - A39rv connectivity | 4.915 | 168 | 0.00000 | 0.00015 | 0.00010 | 0.00020 | A44v - A39c connectivity | 11.423 | 168 | 0.00000 | 0.00907 | 0.00750 | 0.01060 |
| A10m - A40rv connectivity | 9.015 | 168 | 0.00000 | 0.00058 | 0.00050 | 0.00070 | A44v - A39rd connectivity | 15.576 | 168 | 0.00000 | 0.07301 | 0.06380 | 0.08230 |
| A9/46d - A39c connectivity | 10.605 | 168 | 0.00000 | 0.00981 | 0.00800 | 0.01160 | A44v - A40rd connectivity | 22.004 | 168 | 0.00000 | 0.30606 | 0.27860 | 0.33350 |
| A9/46d - A39rd connectivity | 14.952 | 168 | 0.00000 | 0.02385 | 0.02070 | 0.02700 | A44v - A40c connectivity | 23.395 | 168 | 0.00000 | 0.27195 | 0.24900 | 0.29490 |
| A9/46d - A40rd connectivity | 14.638 | 168 | 0.00000 | 0.01142 | 0.00990 | 0.01300 | A44v - A39rv connectivity | 14.737 | 168 | 0.00000 | 0.03730 | 0.03230 | 0.04230 |
| A9/46d - A40c connectivity | 14.926 | 168 | 0.00000 | 0.01353 | 0.01170 | 0.01530 | A44v - A40rv connectivity | 26.387 | 168 | 0.00000 | 0.33392 | 0.30890 | 0.35890 |
| A9/46d - A39rv connectivity | 12.791 | 168 | 0.00000 | 0.00832 | 0.00700 | 0.00960 | A14m - A39c connectivity | 1.408 | 168 | 0.16100 | 0.00001 | 0.00000 | 0.00000 |
| A9/46d - A40rv connectivity | 15.328 | 168 | 0.00000 | 0.00771 | 0.00630 | 0.00810 | A14m - A39rd connectivity | 2.017 | 168 | 0.04500 | 0.00002 | 0.00000 | 0.00000 |
| IFJ - A39c connectivity | 17.811 | 168 | 0.00000 | 0.11677 | 0.10380 | 0.12970 | A14m - A40rd connectivity | 2.133 | 168 | 0.03400 | 0.00002 | 0.00000 | 0.00010 |
| IFJ - A39rd connectivity | 24.869 | 168 | 0.00000 | 0.60423 | 0.55630 | 0.65220 | A14m - A40c connectivity | 3.034 | 168 | 0.00300 | 0.00004 | 0.00000 | 0.00010 |
| IFJ - A40rd connectivity | 23.969 | 168 | 0.00000 | 0.55418 | 0.50850 | 0.59980 | A14m - A39rv connectivity | 1.417 | 168 | 0.15800 | 0.00000 | 0.00000 | 0.00000 |
| IFJ - A40c connectivity | 27.039 | 168 | 0.00000 | 0.52864 | 0.49000 | 0.56720 | A14m - A40rv connectivity | 2.253 | 168 | 0.02600 | 0.00002 | 0.00000 | 0.00000 |
| IFJ - A39rv connectivity | 23.339 | 168 | 0.00000 | 0.22777 | 0.20850 | 0.24700 | A12/47o - A39c connectivity | 2.159 | 168 | 0.03200 | 0.00002 | 0.00000 | 0.00000 |
| IFJ - A40rv connectivity | 16.938 | 168 | 0.00000 | 0.13050 | 0.11530 | 0.14570 | A12/47o - A39rd connectivity | 5.039 | 168 | 0.00000 | 0.00013 | 0.00010 | 0.00020 |
| A46 - A39c connectivity | 7.936 | 168 | 0.00000 | 0.00562 | 0.00420 | 0.00700 | A12/47o - A40rd connectivity | 6.512 | 168 | 0.00000 | 0.00046 | 0.00030 | 0.00060 |
| A46 - A39rd connectivity | 10.83 | 168 | 0.00000 | 0.01748 | 0.01430 | 0.02070 | A12/47o - A40c connectivity | 7.355 | 168 | 0.00000 | 0.00049 | 0.00040 | 0.00060 |
| A46 - A40rd connectivity | 14.769 | 168 | 0.00000 | 0.01101 | 0.00950 | 0.01250 | A12/47o - A39rv connectivity | 5.184 | 168 | 0.00000 | 0.00008 | 0.00000 | 0.00010 |
| A46 - A40c connectivity | 14.986 | 168 | 0.00000 | 0.01372 | 0.01190 | 0.01550 | A12/47o - A40rv connectivity | 7.643 | 168 | 0.00000 | 0.00100 | 0.00070 | 0.00130 |
| A46 - A39rv connectivity | 10.557 | 168 | 0.00000 | 0.00662 | 0.00540 | 0.00790 | A11l - A39rd connectivity | 1.681 | 168 | 0.09500 | 0.00002 | 0.00000 | 0.00010 |
| A46 - A40rv connectivity | 15.582 | 168 | 0.00000 | 0.00749 | 0.00650 | 0.00840 | A11l - A40rd connectivity | 2.698 | 168 | 0.00800 | 0.00006 | 0.00000 | 0.00010 |
| A9/46v - A39c connectivity | 10.421 | 168 | 0.00000 | 0.01972 | 0.01600 | 0.02350 | A11l - A40c connectivity | 2.32 | 168 | 0.02200 | 0.00003 | 0.00000 | 0.00010 |
| A9/46v - A39rd connectivity | 15.857 | 168 | 0.00000 | 0.08146 | 0.07130 | 0.09160 | A11l - A39rv connectivity | 2.683 | 168 | 0.00800 | 0.00002 | 0.00000 | 0.00000 |
| A9/46v - A40rd connectivity | 21.275 | 168 | 0.00000 | 0.08776 | 0.07960 | 0.09590 | A11l - A40rv connectivity | 3.861 | 168 | 0.00000 | 0.00010 | 0.00000 | 0.00010 |
| A9/46v - A40c connectivity | 19.83 | 168 | 0.00000 | 0.10612 | 0.09560 | 0.11670 | A11m - A39rd connectivity | 1.409 | 168 | 0.16100 | 0.00001 | 0.00000 | 0.00000 |
| A9/46v - A39rv connectivity | 14.68 | 168 | 0.00000 | 0.03799 | 0.03290 | 0.04310 | A11m - A40rd connectivity | 1 | 168 | 0.31900 | 0.00000 | 0.00000 | 0.00000 |
| A9/46v - A40rv connectivity | 21.038 | 168 | 0.00000 | 0.06366 | 0.05770 | 0.06960 | A11m - A40c connectivity | 1.411 | 168 | 0.16000 | 0.00001 | 0.00000 | 0.00000 |
| A8vl - A39c connectivity | 13.505 | 168 | 0.00000 | 0.07686 | 0.06560 | 0.08810 | A11m - A39rv connectivity | 1 | 168 | 0.31900 | 0.00000 | 0.00000 | 0.00000 |
| A8vl - A39rd connectivity | 17.006 | 168 | 0.00000 | 0.22301 | 0.19710 | 0.24890 | A11m - A40rv connectivity | 1 | 168 | 0.31900 | 0.00000 | 0.00000 | 0.00000 |
| A8vl - A40rd connectivity | 15.052 | 168 | 0.00000 | 0.09471 | 0.08230 | 0.10710 | A13 - A39c connectivity | 1 | 168 | 0.31900 | 0.00000 | 0.00000 | 0.00000 |
| A8vl - A40c connectivity | 17.535 | 168 | 0.00000 | 0.09849 | 0.08740 | 0.10960 | A13 - A39rd connectivity | 1.418 | 168 | 0.15800 | 0.00001 | 0.00000 | 0.00000 |
| A8vl - A39rv connectivity | 15.741 | 168 | 0.00000 | 0.06910 | 0.06040 | 0.07780 | A13 - A40rd connectivity | 1.739 | 168 | 0.08400 | 0.00001 | 0.00000 | 0.00000 |
| A8vl - A40rv connectivity | 16.692 | 168 | 0.00000 | 0.02542 | 0.02240 | 0.02840 | A13 - A40c connectivity | 1.74 | 168 | 0.08400 | 0.00001 | 0.00000 | 0.00000 |
| A10l - A39c connectivity | 4.295 | 168 | 0.00000 | 0.00100 | 0.00050 | 0.00150 | A13 - A39rv connectivity | 1 | 168 | 0.31900 | 0.00000 | 0.00000 | 0.00000 |
| A10l - A39rd connectivity | 8.329 | 168 | 0.00000 | 0.00342 | 0.00260 | 0.00420 | A13 - A40rv connectivity | 2.013 | 168 | 0.04600 | 0.00001 | 0.00000 | 0.00000 |
| A10l - A40rd connectivity | 10.324 | 168 | 0.00000 | 0.00323 | 0.00260 | 0.00390 | A12/47l - A39c connectivity | 4.816 | 168 | 0.00000 | 0.00017 | 0.00010 | 0.00020 |
| A10l - A40c connectivity | 9.589 | 168 | 0.00000 | 0.00382 | 0.00300 | 0.00460 | A12/47l - A39rd connectivity | 7.739 | 168 | 0.00000 | 0.00084 | 0.00060 | 0.00110 |
| A10l - A39rv connectivity | 6.816 | 168 | 0.00000 | 0.00145 | 0.00100 | 0.00190 | A12/47l - A40rd connectivity | 10.231 | 168 | 0.00000 | 0.00249 | 0.00200 | 0.00300 |
| A10l - A40rv connectivity | 10.881 | 168 | 0.00000 | 0.00296 | 0.00240 | 0.00350 | A12/47l - A40c connectivity | 8.653 | 168 | 0.00000 | 0.00363 | 0.00280 | 0.00450 |
| A44d - A39c connectivity | 12.513 | 168 | 0.00000 | 0.00696 | 0.00590 | 0.00810 | A12/47l - A39rv connectivity | 7.104 | 168 | 0.00000 | 0.00049 | 0.00040 | 0.00060 |
| A44d - A39rd connectivity | 14.348 | 168 | 0.00000 | 0.05458 | 0.04710 | 0.06210 | A12/47l - A40rv connectivity | 11.218 | 168 | 0.00000 | 0.00610 | 0.00500 | 0.00720 |

**Supplementary Table 2.** One sample t-test results of the connectivity between the selected subregions within the PFC-IPC.

| Supplementary Table 3. Track density percentile for the PFC subregions regarding the PFC-TC connectivity | | | |
|---|---|---|---|
| PFC Subregions | Total Track Density for the Reached Streamlines | Percentage (%) | Percentile (%ile) |
| A8m | 0.212910494 | 0.78 | 45.4 |
| A8dl | 0.509150703 | 1.87 | 59 |
| A9l | 0.163046633 | 0.60 | 40.9 |
| A9m | 0.063010392 | 0.23 | 27.2 |
| A10m | 0.022251054 | 0.08 | 22.7 |
| A9/46d | 0.252953499 | 0.93 | 50 |
| IFJ | 7.28754887 | 26.73 | 100 |
| A46 | 0.268273183 | 0.98 | 54.5 |
| A9/46v | 2.437447081 | 8.94 | 81.8 |
| A8vl | 1.207515367 | 4.43 | 72.7 |
| A10l | 0.088954885 | 0.33 | 31.8 |
| A44d | 2.621834148 | 9.61 | 86.3 |
| IFS | 1.986949193 | 7.29 | 77.2 |
| A45c | 3.856911598 | 14.14 | 90.9 |
| A45r | 0.520196853 | 1.91 | 63.6 |
| A44op | 1.085777691 | 3.98 | 68.1 |
| A44v | 4.533456365 | 16.63 | 95.4 |
| A14m | 0.001121574 | 0.00 | 9 |
| A12/47o | 0.020195708 | 0.07 | 18.1 |
| A11l | 0.001861091 | 0.01 | 13.6 |
| A11m | 1.59651E-05 | 0.00 | 0 |
| A13 | 0.000518674 | 0.00 | 4.5 |
| A12/47l | 0.126536029 | 0.46 | 36.3 |

**Supplementary Table 3.** Track density percentile for the PFC subregions regarding the PFC-TC connectivity.

**Supplementary Table 4.** One sample t-test results of the connectivity between the selected subregions within the PFC-TC.

**Supplementary Table 5.** One sample t-test results of the connectivity between the selected subregions within the PFC.

| Track Density | T | df | Sig. (2-tailed) | Mean Difference | 95% Confidence Interval of the Difference Lower | 95% Confidence Interval of the Difference Upper | Track Density | T | df | Sig. (2-tailed) | Mean Difference | 95% Confidence Interval of the Difference Lower | 95% Confidence Interval of the Difference Upper | Track Density | T | df | Sig. (2-tailed) | Mean Difference | 95% Confidence Interval of the Difference Lower | 95% Confidence Interval of the Difference Upper | Track Density | T | df | Sig. (2-tailed) | Mean Difference | 95% Confidence Interval of the Difference Lower | 95% Confidence Interval of the Difference Upper |
|---|---|---|---|---|---|---|---|---|---|---|---|---|---|---|---|---|---|---|---|---|---|---|---|---|---|---|---|

**Supplementary Table 5.** One sample t-test results of the connectivity between the selected subregions within the PFC.